\documentstyle[12pt]{article}
\newcommand{\be}{\begin{equation}}
\newcommand{\ee}{\end{equation}}
\newcommand{\nn}{\nonumber\\}
\newcommand{\Dc}{{\bf D}}
\newcommand{\dtp}{\tilde{\tilde{\phi}} }
\newcommand{\ga}{\alpha }
\newcommand{\ec}{et\,cycl.}

\newcounter{mycount}

\newcommand{\bee}{\begin{eqnarray}}
\newcommand{\eee}{\end{eqnarray}}
\newcommand{\D}{{\cal D}}
\newcommand\lr {\leftrightarrow}

\newcommand\lra {\leftrightarrow}

\newcommand\half{\frac 1 2 }

\begin{document}

\newcommand\go{\omega}

\vspace{-1mm}
\begin{flushright} G\"{o}teborg ITP 95-20 \\
\vspace{-1mm}
 FIAN/TD/16--95\\
\vspace{-1mm}
{September 1995}\\
\end{flushright}\vspace{2cm}

\begin{center}
{\Large{\bf Generalized Toda Field Theories }}
\vspace{15mm}\\
{\bf\large Lars ~Brink }\vspace{5mm}\\
Institute for Theoretical Physics, S-412 96 G\"{o}teborg,
Sweden\vspace{1.5cm}\\
{\bf\large Mikhail A.~Vasiliev}
\vspace{0.5cm}\\
I.E.~Tamm Theoretical Department, P.N.~Lebedev
Physical Institute, \\117924, Leninsky Prospect 53, Moscow, Russia

\begin{abstract}
In this paper we introduce a unified approach to Toda field theories
which allows us to formulate the classes of $A_n$, $B_n$ and $C_n$ models
as unique models involving an arbitrary
continuous parameter $\nu$. For certain values of $\nu $,
the model describes the standard Toda theories.
For other values of $\nu$ it defines a class of models that involve
infinitely many fields.
These models interpolate between the various standard
Toda field theories. They are conformally invariant and possess infinitely
many conserved higher-spin currents thus making them
candidates for a new set of integrable systems.
A general construction
is performed, which can effectively be used for the
derivation of explicit forms of particular higher-spin currents.
We also study the currents in a different representation
in which they are linear in the dynamical variables
having, however, a non-linear Poisson bracket algebra.
An explicit formula for this Poisson structure is found.

\end{abstract}
\end{center}

\newpage

\noindent

\vskip 2mm

\section{Introduction}

Toda field theory encodes a class of 2d field theories
which are both integrable \cite{int1,int2,int3} and conformal
\cite{conf1,conf2,conf3} and therefore
quite interesting from various points of view.
These models have attracted a considerable interest
in recent years both from a theoretical point of view
as  well as from a physical one
 (for a review see e.g. \cite{rev} and references therein).
The particular Toda models are classified according to the
classical
Lie algebra root systems  and form thus an infinite set
of models. In this paper we introduce a unified approach which
allows us to formulate the classes of $A_n$, $B_n$ and $C_n$ models
as unique models involving an arbitrary
continuous parameter $\nu$. For $2\nu =2n+1$, where
$n$ is a non-negative integer, the model describes
 the standard $A_n$, $B_n$ and $C_n$ (or $A_{n-1}$, $B_{n-1}$
and $C_{n-1}$) Toda theories. For $2\nu\neq 2n+1$
it defines a class of models that involve infinitely many fields.
These models interpolate between the various standard (finite)
Toda field theories \cite{root}. They
are also conformally invariant and possess infinitely
many conserved higher-spin currents thus making them
candidates for a new set of integrable systems. The limiting
case $\nu \rightarrow \infty$ corresponds to the continuous
Toda system \cite{cont}.

The scheme we will follow is based on infinite-dimensional algebras
proposed in \cite{s2} in the context of the analysis of higher-spin
theories. These algebras described in Section 2
belong to a class of algebras denoted as
$W_{1+\infty}$ (or wedge algebras) \cite{Win}.
The proposed model turns out
indeed to be deeply related to higher-spins physics. It corresponds to vacuum
solutions of the class of 2d higher-spin models introduced recently in
\cite{2d} as well as to dimensionally reduced solutions of the $d=3$
\cite{un,d3} and $d=4$ \cite{more} higher-spin gauge theories.
{}From this viewpoint the Toda systems we focus on in this paper are
of interest as describing various background geometries of the higher-spin
gauge models. Note that this interpretation is parallel to the
interpretation of the ordinary Toda field theories as particular solutions
of the $W$ - gravity models described e.g. in \cite{wgr}.

The unified character of our approach allows us to study
properties of the model in different representations. In
one such representation described in Section 3 the equivalence
to the ordinary Toda field theories can be established.
This is done in Section 4. The higher-spin symmetries are discussed
in Section 5 and the general construction of the corresponding currents
is performed in Section 6.
In that section we also study the currents in a different representation
in which they are linear in the dynamical variables
having, however, a non-linear Poisson bracket algebra. In this representation
the study of the algebra of the currents is much simplified.
The form of this algebra seems to be new and is different from that of the
algebra $\tilde{W}_{1+\infty}$ considered in \cite{tw}.
This method can effectively be used for the
derivation of explicit forms of particular local higher-spin currents
both in infinite and finite Toda field theories in which case
the problem of constructing local higher-spin currents  has been
extensively studied by
various methods (see e.g. \cite{int2}-\cite{rev}, \cite{scur,exp}).
We demonstrate this for some simplest examples in Section 7 where
we also analyze conformal properties of higher-spin
currents. Appendices A and B contain some technical details of
the proofs.
The main definitions and notations are collected in Appendix C.

In this paper we confine ourselves to the study of the classical
models.

\setcounter{equation}{0}

\section{Algebra}

The central algebraic element of our approach is to
consider a
function $\phi (T^0 )$ of the operator $T^0$,
instead of
describing dynamical fields of the Toda system $\phi^i$ which take
their values in the Cartan subalgebra of $A_n$, where
\be
\label{T0}
T^0 = \frac{1}{4} \{a^- , a^+ \}\,.
\ee
The oscillators $a^\pm$  obey the
commutation relation
\be
\label{alg}
[a^- , a^+ ] =1 +2\nu K \,,
\ee
where $\nu$ is an arbitrary constant while the operator $K$ possesses
the properties
\be
\label{K}
Ka^\pm =-a^\pm K\,;\qquad K^2 =1 \,.
\ee

The crucial property of the commutation relations (\ref{alg}), (\ref{K}) is
that the operator $T^0$ along with
the generators
\begin{equation}
\label{Tpm}
T^\pm = \frac{1}{2} (a^\pm )^2
\end{equation}
form $sl_2$ commutation relations
\begin{equation}
\label{sl2}
[T^0 ,T^\pm ] =\pm T^\pm\,,\qquad [T^- ,T^+ ]=2T^0
\end{equation}
independently of the value of $\nu$. Note that based on the definitions
(\ref{T0})-(\ref{Tpm}), one can derive the following useful formulae
\begin{equation}
\label{TT}
T^- T^+ =(T^0 )^2 + T^0 - C\,,\qquad
T^+ T^- =(T^0 )^2 - T^0 - C\,,
\end{equation}
where
\begin{equation}
\label{C}
C =-\frac{1}{16} (3 +4 \nu K -4 \nu^2 )
\end{equation}
is the quadratic Casimir operator of $sl_2$ in the realization
(\ref{T0})-(\ref{Tpm}).
Since $K$ commutes with even combinations of the oscillators
$a^\pm$, it can be replaced with 1 when acting on even
functions of $a^\pm$  (the value
$-1$ is equivalent by $\nu\lr -\nu$). As a result, this
realization allows one to describe various types of representations
of $sl_2$ corresponding to different values of the Casimir operator
in a uniform way.

Let us note that the modified oscillators (\ref{alg}) have been
considered in the literature in various contexts and realizations
\cite{osc}. In \cite{s2} the relations (\ref{alg}),(\ref{K}) were considered
as the defining relations for an associative algebra $Aq(2;2\nu )$ with the
generating elements $a^\pm$ and $K$. This algebra can be shown
\cite{BVW} to
coincide with the factor of the enveloping algebra of $osp(1;2)$,
$U(osp(1;2))$, over
its ideal generated by the quadratic Casimir operator
by factoring out all elements
of the form $(C_2 - c_2)a$ where $a\in U(osp(1;2))$ and
$c_2=\frac{1}{16}(4\nu^2 -1)$ is an arbitrary constant.
The associative algebras $Aq (2;2\nu )$ give rise to the Lie
(super)algebras with the (super)commutator $[f,g\}$ as a product law.
These algebras are known in the literature as
centerless (super) $W_\infty (\lambda )$ \cite{BVW}
($\lambda = \frac{1}{4} (2\nu +1 )$) or wedge algebras
\cite{Win}.
Note that the simple components of the even subalgebra of $Aq(2;2\nu )$,
which are isomorphic to the factor algebras of $U(sl_2 )$
over appropriate ideals generated by the quadratic Casimir
operator, were studied in \cite{SH}.
Infinite-dimensional algebras of this class play a fundamental
role in the higher-spin gauge theories developed in
\cite{HS}, \cite{un,d3,more} (and references therein). In \cite{d3} it was
argued in particular that $\nu$ can be interpreted as a vacuum value of a
certain scalar field in the theory of higher-spin-matter interactions in three
space-time dimensions.

A general element of $Aq(2;2\nu )$ can be cast into the form
\begin{equation}
\label{gel}
B = \sum_{A=0}^1 \sum_{n=0}^\infty \frac{1}{n!}
b^A_{\, \alpha_1 \ldots \alpha_n } K^A\,
a^{\alpha_1} \ldots a^{\alpha_n }\,,
\end{equation}
where the coefficients $b^A_{\,\alpha_1 \ldots \alpha_n }$
are totally symmetric in the (spinorial) indices $\alpha_n$ ($\alpha =\pm$)
which implies Weyl ordering. Note that the indices
are raised and lowered by the
$2\times 2$ symplectic form $\epsilon_{\alpha\beta}$=$-\epsilon_{\beta\alpha}$,
$\epsilon_{-+}=\epsilon^{-+}=1$, $A_\alpha$=$A^\beta \epsilon_{\beta\alpha}$,
 $A^\alpha$=$ \epsilon^{\alpha\beta}$ $A_\beta$.

In the sequel we will use the following notation for the
basis elements of the algebra
\be
\label{base}
E_{n,m}=\Big ( (a^+ )^n (a^- )^m \Big )_W \equiv\frac{1}{(n+m)!}
\Bigl ( (a^+ )^n (a^- )^m +((n+m)!-1)  )\,\mbox{permutations}\Bigr ) \,.
\ee
Equivalently one can say that $E_{n,m} $ is such an element $B$
(\ref{gel}) in which $b^{A}_{ \alpha_1 \ldots \alpha_{n+m} }$ is different
from zero only when $A=0$  and $n$ out of $n+m$ indices $\alpha$ take
the value $``+"$
while the other $m$ indices take the value $``-"$.
We will use the notation $h_n$ for the grade-zero elements $E_{n,n}$ as they
form a basis of the Cartan-type commutative subalgebra of $Aq(2;2\nu )$
\be
\label{wn}
h_n =E_{n,n}\,.
\ee
Note that
\be
\label{wsl2}
T^+ = \frac{1}{2}E_{2,0}\,,\qquad
T^- = \frac{1}{2}E_{0,2}\,,\qquad
T^0 = \frac{1}{2}h_1 \,.
\ee

It should be mentioned that there is no proper generalization
of the Cartan-Chevalley basis to the full infinite-dimensional
algebra $Aq (2;2\nu )$. Note however that a more general
concept of ``continuum Lie algebras" \cite{con} might be useful in this
respect.

It is an important property of $Aq(2;2\nu )$ that it possesses a
supertrace operation \cite{s2}, i.e. a linear complex valued
function $str()$ on $Aq(2;2\nu )$ such that
\begin{equation}
\label{strprop}
str (AB) =
(-1)^{\pi(A)}str (BA) =
(-1)^{\pi(B)}str (BA)\,,
\end{equation}
where $\pi(B)$
is defined to be equal to $0$ or $1$ for monomials with
even or odd powers $n$ in (\ref{gel}), respectively.

In \cite{s2} it was shown that there exists a unique
supertrace operation
(up to a scale factor)
on $Aq(2;2\nu)$
\begin{equation}
\label{str}
str (B) = b^0 -2\nu b^1 \,.
\end{equation}

The existence of a (super)trace operation is important in many
respects. In particular, the (super)trace allows one to construct
invariant forms on the algebra as
\be
\label{forms}
str(B_1 B_2 \ldots B_n )\,.
\ee

As a result one can construct invariants essentially in the same way as
for finite-dimensional algebras with the aid of an ordinary (super)trace.
{}From the viewpoint of field theoretical applications this gives a
possibility to construct invariant Lagrangians with the field variables
taking their values in $Aq(2;2\nu )$.
Another useful application of invariant forms is that null vectors of any
invariant bilinear form span some ideal of the algebra. As a result,
having a supertrace operation one has a constructive possibility to
search for ideals of the algebra. Moreover, the ideals of this particular
class have the very important property that they automatically decouple from
the factor algebra under the supertrace operation.

In \cite{s2} an explicit formula for the bilinear invariant form
induced by the supertrace (\ref{str}) was derived. The final result
is
\be
\label{bil}
str(B_1 B_2 )=\sum_{n=0}^\infty \sum_{A,B =0,1} \frac{1}{n!}
\beta^{AB} (n) b^A_{1\, \alpha_1 \ldots \alpha_n }
b_2^{B\, \alpha_1 \ldots \alpha_n }\,,
\ee
where
\bee
\label{be}
\beta^{AB} (n)&=&(-1)^{Bn} 2^{-n}\prod_{l=0}^{[(n-1)/2]}
\bigl(1-\frac{4\nu^2}{(2l+1)^2}\bigr)\times\nonumber\\
&{}&\bigl[ \delta (A+B) -\frac{1}{2}(1+(-1)^n )
\frac{2\nu}{n+1}\delta (A+B+1)\bigr]
\eee
for $n> 0$ and
\be
\label{be0}
\beta^{AB} (0)=
[ \delta (A+B) -
2\nu\delta (A+B+1)]\,.
\ee
Here $\bigl[ \frac{m}{2}\bigr]$ denotes the integral part of
$\frac{m}{2}$ $\bigl(\bigl[ \frac{m}{2}\bigr]\leq \frac{m}{2})$,
while $A+B$ and $A+B+1$ are assumed to take values $0$ or $1$
(i.e. with the convention $1+1=0$).

{}From this formula it follows that
\bee
\label{bilw}
str(K^A E_{n,m}E_{k,l} )=0\quad \mbox{if $n\neq l$ or $m\neq k$}\,,\nn
str(K^A E_{n,m}E_{m,n}) =(-1)^n n!\,m!\beta^{A0} (n+m)\,.
\eee

In this paper we only consider the even subalgebra
of $Aq(2;2\nu )$ spanned by even power polynomials in $a^\pm$
which in its turn decomposes
into a direct sum of two
subalgebras $Aq^\pm (2;2\nu )$ projected out by the projectors
\be
\label{Ppm}
P^\pm =\frac{1}{2} (1\pm K)\,.
\ee

The corresponding bilinear forms read
\bee
\label{bilpm}
str(P^\pm E_{n,m}E_{m,n}) =(-1)^n n!\,m!\beta^\pm (n+m)\,,
\eee
where
\bee
\label{bpm}
\beta^{\pm} (n)&=&  2^{-(n+1)}
\bigl[ 1 \mp \frac{1}{2}(1+(-1)^n )
\frac{2\nu}{n+1}\bigr]\,
\prod_{p=0}^{[(n-1)/2]}
\bigl(1-\frac{4\nu^2}{(2p+1)^2}\bigr)\,.
\eee
For the case of even $n$ under investigation this formula can
be equivalently rewritten as
\bee
\label{bev}
\beta^{\pm} (2m)&=&  2^{-(2m+1)}\prod_{p=-m}^{p=m}
\bigl(1\mp \frac{2\nu}{2p+1}\bigr) \,.
\eee

{}From this formula it follows that the bilinear form $str(AB)$
degenerates for any value of $2\nu =2l+1$, where
$l$ is an arbitrary integer. As a result, all $P^+ E_{n,m}$ with
$n+m \geq 2l$ and $P^- E_{n,m}$ with
$n+m \geq 2(l+1)$
decouple from everything under the supertrace
operation. This means that all these higher-order polynomials
in $a^\pm$ form an ideal $I^\pm_l$ of $Aq^\pm (2l+1 )$ which can be factored
out (and in fact factors out automatically under the supertrace
operation). As a result one is left with the finite-dimensional
factor algebra of lower-order polynomials. Its dimensionality is
$l^2$ and $(l+1)^2 $ in the case of $Aq^+ (2l+1)$
and $Aq^- (2l+1)$, respectively \cite{s2}.
Since this finite-dimensional algebra possesses a non-degenerate
bilinear form, one concludes (see e.g. \cite{ass})
that it is isomorphic to the
matrix algebras $Mat_{l}$ and $Mat_{l+1}$, respectively, i.e.
\be
\label{fact}
Aq^+ (2l+1 )/ I^+_l \,=Mat_l\,\qquad
Aq^- (2l+1 )/ I^-_l =Mat_{(l+1)}\,.
\ee

Let us note that from the basic definition of the supertrace
(\ref{str}) it follows that for $2\nu =2l+1$
\be
\label{mattr}
str(P^+ )=- l\,,\qquad str(P^- )=l+1\,,
\ee
which agrees with the standard definition of the trace of matrix algebras,
$tr (1)=n$ for $Mat_n$.
The negative relative sign in the relations (\ref{mattr})
is due to the invariance of the full original algebra with respect
to $\nu\lra -\nu$, $K\lra -K$ which is equivalent to $P^\pm \lra P^\mp$,
$l\lra -(l+1)$.

Taking into account these results we introduce the notation $a(\nu \mp\half )$
for the Lie algebras of commutators originating from  $Aq^\pm (2\nu \mp 1 )$

In accordance with the standard definition,
a linear
mapping $\rho$ of an algebra $A$ onto itself such that
\be
\label{anti}
\rho (ab)=\rho(b)\rho(a)\,,\qquad \forall a,b \in A\,.
\ee
is called an antiautomorphism of $A$.
The algebra $Aq(2;2\nu )$ admits \cite{s2} the following antiautomorphism
$\rho (P^\pm E_{n,m})
=i^{n+m}P^\pm  E_{n,m}$.
To make sure that the property (\ref{anti}) is true it is enough to
check that it is in agreement with the defining relations (\ref{alg}) and
(\ref{K}). Obviously, any antiautomorphism of $Aq(2;2\nu )$ induces some
automorphism of the related $a(\nu\mp\half )$ algebras of commutators simply by
changing a sign. An automorphism $\tau$ related in this way to $\rho$ therefore
is
\be
\label{tau}
\tau (P^\pm E_{n,m})
=-i^{n+m}P^\pm  E_{n,m}\,.
\ee
By using the automorphism $\tau$ one can project out subsystems of
elements satisfying the condition
\be
\label{trunc}
\tau (a)=a\,,
\ee
which span a proper subalgebra of the original $a(\nu\mp\half )$ Lie algebras.

For the case of $2\nu =2l+1$ with $l=2n$ one finds \cite{s2} that the
respective
truncated subalgebras of
$Aq^+ (l+\half )/ I^+_l$ is $sp(2n)\sim c_n$ and of
$Aq^- (l+\half )/ I^-_l$ is $o(2n+1)\sim b_n $.
For $l=2n+1$ one gets $sp(2n+2)\sim c_{n+1}$ as
a truncation of
$Aq^+ (l+\half )/ I^+_l$
and $o(2n+1)\sim b_n$ as
a truncation of
$Aq^- (l+\half )/ I^-_l$. Note that it is impossible to extract the $o(2n)=d_n$
algebra in this way.

It is important to note that the whole analysis below turns out to
be invariant under the action of the automorphism $\tau$. This means
that the case of algebras $a(\sigma )$ considered in the sequel of this
paper can be trivially truncated to the $b(\sigma )$ and $c(\sigma )$ - type
Toda field theories.

\setcounter{equation}{0}

\section{Generalized Toda Dynamics}

The infinite-dimensional Lie algebras $a(\mu )$
have been used
as the gauge symmetry algebras in the construction of higher-spin
gauge theories in three and two space-time dimensions \cite{2d,d3}.
The $2d$ gauge fields of this algebra can be written as
\begin{equation}
\label{Gauge}
W_\mu = \sum_{n=0}^\infty \frac{1}{n!}
W_\mu^{A}{}_{ \alpha_1 \ldots \alpha_n } (z)
K^A\, a^{\alpha_1} \ldots a^{\alpha_n }\,,
\end{equation}
where $W_\mu^{A}{}_{\alpha_1 \ldots \alpha_n } (z)$
are the higher-spin gauge fields,
$z^\mu $ denotes the two-dimensional space-time coordinates
and $\mu =\pm $.

The curvatures have the standard form
\begin{equation}
\label{cur}
R_{\nu\mu} = \partial_\nu W_\mu - \partial_\mu W_\nu +
[W_\nu ,W_\mu ]\,.
\end{equation}
where $\partial_\pm$=$\frac{1}{\sqrt{2}}(\partial_0 \pm \partial_1 )$.

Setting the curvature to zero allows us to find vacuum solutions
of the higher-spin gauge theory. The simplest vacuum solution which
describes the
anti-de Sitter vacuum geometry is obtained by constraining
$W_\mu $ to be of the form
\begin{equation}
\label{ads}
W_\mu = h_{\mu}^+ (z) T^+ + h_{\mu}^- (z) T^- +\omega_\mu (z) T^0 \,,
\end{equation}
where $h_\mu^\pm$ are
zweibein one-forms and $\omega_\mu$ is the Lorentz connection
one-form. This ansatz is consistent because the generators
$T^\pm$ and $T^0$ span a proper $sl_2$ subalgebra of the full algebra.
The components of the $sl_2$ curvatures then read
\be
\label{sl2c}
R_{\nu\mu} = R_{\nu\mu}^+ T^+ + R_{\nu\mu}^- T^- + R_{\nu\mu}^0 T^0 \,,
\ee
with
\be
R_{\nu\mu}^\pm = \partial_\nu h^\pm_\mu  \pm
\omega_\nu h^\pm_\mu -(\nu\lra\mu )\,,
\ee
\be
R_{\nu\mu}^0 = \partial_\nu \omega_\mu  +
2h^-_\nu h^+_\mu -(\nu\lra\mu )\,.
\ee

By using a conformal gauge
\be
\label{cg}
h_\mu^\pm =\delta_\mu^\pm exp(\varphi^\pm )\,,
\ee
the zero-curvature conditions
$R_{\nu\mu}^\pm=0$ allows us to solve for $\omega_\nu$
\begin{equation}
\omega_\pm = \pm \partial_\pm \varphi^\mp\,.
\end{equation}
Inserting this into the equation $R_{\nu\mu}^0=0$ we are left with the
Liouville
equation
\be
\partial_+ \partial_- (\varphi^+ +\varphi^- )
= -2exp [ (\varphi^+ +\varphi^- )]\,.
\ee
Note that the fields $\varphi^\pm$ appear only in the combination
$\varphi^+ +\varphi^- $ since the local Lorentz symmetry still remains unfixed.
One can take any convenient gauge condition like, say,  $\varphi^- =0$.

In order to obtain the generalized Toda system we
will follow the general zero-curvature approach \cite{int1,rev,wgr} and
consider the following constraints for the higher-spin gauge fields
\begin{equation}
\label{ans}
W_\mu = \delta_{\mu}^+ H^+ + \delta_{\mu}^- H^- +\omega_\mu (T^0 )\,,
\end{equation}
where the index $\pm$ indicates that the element $H^\pm$ has grade
$\pm 1$ in the sense that $[T^0 ,H^\pm ]$ =$ \pm H^\pm $ while
$[T^0 ,\omega_\mu (T^0 )]=0$.
The
non-trivial curvatures can then be written as
\begin{equation}
\label{R}
R_{\nu\mu} = R_{\nu\mu}^+ + R_{\nu\mu}^- + R_{\nu\mu}^0\,,
\end{equation}
where
\begin{equation}
R_{\nu\mu}^\pm =  \delta_{\mu}^\pm \partial_\nu H^\pm -
\delta_{\nu}^\pm \partial_\mu H^\pm +
\delta_\mu^\pm[\omega_\nu , H^\pm ] -\delta_\nu^\pm[\omega_\mu , H^\pm ]\,,
\end{equation}
\begin{equation}
R_{\nu\mu}^0 = \partial_\nu \omega_\mu
+\delta_\nu^+ \delta_\mu^- [H^+ ,H^- ]
-\nu\leftrightarrow \mu\,.
\end{equation}

A convenient parametrization of $H^\pm$ is
\be
\label{hpm}
H^+ =e^{\phi_+} T^+ e^{-\phi_+ }\,,\qquad
H^- =e^{-\phi_-} T^- e^{\phi_-} \,,
\ee
where $\phi_\pm$ are general functions of
the generator $T_0$ and of $K$.

Imposing the zero-curvature conditions
 $R^\pm_{\nu\mu} =0 $  one finds
\begin{equation}
\label{connec}
\omega_\pm = \pm \partial_\pm \phi_\mp\,,
\end{equation}
while from $R^0 =0$ one gets
\begin{equation}
\label{equ}
\partial_- \partial_+ (\phi_+ +\phi_- ) = [H^+ ,H^- ]\,.
\end{equation}

Again we have a Lorentz symmetry-type gauge ambiguity which
can be used to set one of the $\phi_\pm$ fields to zero.
{}From now on we will use the gauge
\be
\label{gauge}
\phi_+=0\,,\rightarrow\omega_- =0\,,\quad H^+ =T^+
\ee
and denote
$\phi_- =\phi$ using the label $\pm$ in a different context
from above to indicate projections by $P^\pm$ (\ref{Ppm}). The general
expansion is then
\be
\label{phw}
\phi^\pm =P^\pm \phi=\sum_{n=0}^\infty P^\pm \phi^\pm_n h_{n}\,,
\ee
where $h_n$ is defined in (\ref{wn}).

In the sequel we will show how these equations
include the $A_n$ Toda equations for the specific values of the
parameter $2\nu = 2n+1$. For other values of $\nu$ the same
equations describe an infinite component system.

Note that by rescaling the $\phi$ fields as
$\phi \rightarrow \beta\phi$ one can rewrite
the equations (\ref{equ}) in the equivalent form
\begin{equation}
\label{eq}
\partial_- \partial_+ \phi  =\beta^{-1}
[T^+ , e^{-\beta\phi} T^- e^{\beta\phi}]\,.
\end{equation}
Since in this paper we only treat the classical case
we will use this freedom to set $\beta$ to one.

When the bilinear form
(\ref{bil}) is non-degenerate, i.e. for all $2\nu\neq 2l+1$,
where $l$ is an integer,
the equations of motion (\ref{equ}) can be elevated to an action
\begin{equation}
\label{action}
S^\pm =-\int d^2 z\, str\Big( P^\pm ( \phi \partial_- \partial_+ \phi
- 2H^+ H^- )\Big ) \,.
\end{equation}
When  $2\nu= 2l+1$, the null-vector ideals do not contribute to
the supertrace. As a result
the action reduces to the
finite-dimensional system automatically.

The action (\ref{action}) is conformally invariant under transformations
of the form
\begin{equation}
\label{conf}
\delta \phi =  u(z^+ )\partial_+ \phi (T^0 ,z) + \partial_+ u(z^+ )T^0\,,
\end{equation}
where $u(z^+ )$ is an arbitrary parameter function.
We will come back to the analysis of the symmetries of the
action (\ref{action}) in section 5.

\setcounter{equation}{0}

\section{Comparison to Toda Field Theories}

In order to make contact for the action (\ref{action})
with the usual Toda systems one first
of all must be able to compute expressions of the form
\be
\label{pot}
V^\pm (\nu ;\phi )=-2str\left (P^\pm  T^+ e^{-\phi} T^- e^{\phi}\right )
\ee
for any $\phi$ of the form (\ref{phw}). This can be done in two
steps. First one moves $T^-$ through the exponential reducing the
problem to compute expressions of the form
\be
\label{gf}
Z^\pm(\nu ;\alpha )=str \left (P^\pm exp \sum_{n}(\alpha_{n}h_{n})
\right )\,.
\ee

The first step can be carried out by virtue of the
following useful formula
\be
\label{shif}
a^- h_{n} =\tilde{h}_{n}a^- \,,
\ee
where
\be
\label{tildew}
\tilde{h}_{n}=
\sum_{l=0}^n   A_l^{n} (\nu ) h_{n-l}
\ee
and
\be
\label{als}
A_l^{n} (\nu ) = \frac{1}{2^{l-1}} \frac{n\,!}{(n-l)\,!}
\prod_{m=1}^{l-1} \left ( 1-\frac{2\nu K}{2n+1-2m}\right )\,;\quad
A_0^n (\nu )=1\,,\quad A_1^n (\nu )=n\,.
\ee

The formula (\ref{shif}) can be derived by using properties
of Weyl ordered polynomials of $a^\pm$.
By using this formula
twice one arrives at a similar expression for the case when
$T^-$ is commuted through $h_{n}$
\be
\label{shiftT}
T^- h_{n} =\tilde{\tilde{h}}_{n}T^- \,.
\ee

With the aid of this formula one obtains
\bee
\label{spot}
str\left (P^\pm T^+ e^{-\phi} T^- e^{\phi}\right )&=&
str\left (P^\pm T^- T^+ e^{(\dtp -\phi )}\right )\nn
&=&\frac{1}{4} str\left (P^\pm \left ((h_1)^2 + 2 h_1 +
\frac{1}{4} (3 \pm 4 \nu  -4 \nu^2 )\right ) e^{(\dtp -\phi )}\right )\,,
\eee
where we have made use of the formula (\ref{TT}).

As a result the potential (\ref{pot})
can be computed from the generating
function
$Z^\pm (\nu ;\alpha )$ (\ref{gf}) in the following way
\bee \label{gf1}
V^\pm (\nu ;\phi ) =-\frac{1}{2} \left (
\frac{\partial^2}{\partial\alpha_1^2} +2 \frac{\partial}{\partial\alpha_1 }
+\frac{1}{4} (3 \pm 4 \nu  -4 \nu^2 ) \right )
Z^\pm (\nu ;\alpha ) \vert_{\alpha_n =\dtp_n (z) -\phi_n (z)}\,.
\eee

Let us now explain how one can compute generating functions of this
form. First, one computes the simplest generating function
\be
\label{gfy}
Y^\pm (\nu ;\alpha )=str \left ( P^\pm exp (\alpha h_{1})\right )\,,
\ee
which can be shown to be
\be
\label{mgf}
Y^{\pm} (\nu ;\alpha )= \frac{1}{2}\left (
\frac{ch(\nu\alpha )}{ch(\frac{\alpha}{2} ) } \mp\frac{sh(\nu\alpha )}{sh(
\frac{\alpha }{2} ) }
\right ) =\mp \frac{sh ((\nu \mp\half )\alpha )}{sh(\alpha )}\,.
\ee

In order to derive this formula one starts by differentiating $Y(\alpha )$
with respect to $\alpha$. This brings down $h_1=\half \{a^- ,a^+ \}$.
By using the properties of the supertrace operation (\ref{strprop}) and the
simplest
version of the formula (\ref{shif})
\be
\label{shift0}
a^-h_1 =(h_1 +1 ) a^-
\ee
one gets
\bee
&{}&\frac{\partial}{\partial \alpha}Y^{\pm} (\nu ;\alpha ) = \half
str\left ( P^\pm a^- a^+ e^{\alpha h_1 } \right )-\half
str\left ( P^\mp a^- a^+ e^{\alpha ( h_1 +1) } \right )\nn
&{}&=\half
str\left ( P^\pm (\frac{\partial}{\partial \alpha} +
\half (1\pm 2\nu ))e^{\alpha h_1 } \right )-\half e^\alpha
str\left ( P^\mp (\frac{\partial}{\partial \alpha} +
\half (1\mp 2\nu )) e^{\alpha h_1 } \right )\,.
\eee
{}From this we read off the following differential equations
with respect to $\alpha$
\be
\label{eqy}
\frac{\partial}{\partial \alpha}Y^{\pm} +e^\alpha
\frac{\partial}{\partial \alpha}Y^{\mp} =\half (1\pm 2\nu )Y^\pm
-\half e^\alpha(1\mp 2\nu )Y^\mp\,.
\ee
The initial data which follow from (\ref{str}) are
\be
\label{bc}
Y^\pm (\nu ; 0)=\half (1\mp 2\nu )\,.
\ee
The formula (\ref{mgf}) is then a unique solution of
equation (\ref{eqy}) under the boundary conditions (\ref{bc}).

Let us now consider the general generating function (\ref{gf}).
It follows that
\be
\label{t1}
\frac{\partial Z^\pm (\nu ;\alpha )} {\partial \alpha_{n+1}}
=str\left (P^\pm h_{n+1}
exp(\sum_n \alpha_n h_n ) \right )\,.
\ee

One can prove the following relation
\be
\label{t2}
P^\pm h_{n+1} =P^\pm h_n h_1 +P^\pm d_\pm^n (\nu ) h_{n-1}\,,
\ee
where
\be
\label{t3}
d_\pm^n (\nu )= \frac{1}{4} n^2 \left ( 1\mp \frac{2\nu }{2n+1 }\right )
\left ( 1\pm \frac{2\nu }{2n-1 }\right )\,.
\ee
The equations (\ref{t2}) and (\ref{t3}) are derived by using the
properties of Weyl ordered polynomials in $a^\pm$ as well as the
commutation relations (\ref{alg}).

It then follows that the generating function (\ref{gf})
satisfies the following set of differential equations
\be
\label{difa}
\left ( \frac{\partial}{\partial \alpha_{n+1}}-
\frac{\partial}{\partial \alpha_{1}}\frac{\partial}{\partial \alpha_{n}}
-d_\pm^n (\nu )\frac{\partial}{\partial \alpha_{n-1}}\right )
Z^\pm (\nu ;\alpha )=0\,.
\ee

This equation can then be used in an iterative way to solve for
$Z^\pm (\nu ;\alpha )$
once one has solved for the
$Z^\pm (\nu ;\alpha_0 ,\alpha_1 ,0,0\ldots )$=
$e^{\alpha_0} Y^\pm (\nu ,\alpha_1 )$
using
$Z^\pm (\nu ;\alpha_0 ,\alpha_1 ,\ldots ,\alpha_n , 0, \ldots  )$
successively as initial data.
For example, when the initial data are of the form
\be
Z^\pm (\nu ;\alpha_0 ,\alpha_1 ,0,0\ldots )=
e^{\alpha_0} \sum_k c_k e^{u^1_k \alpha_1 }
\ee
with some coefficients $c_k$ and $u^1_k$, then the solution of
(\ref{difa}) reads
\be
Z^\pm (\nu ;\alpha )=\sum_k c_k exp(\alpha_0 +\sum_l u^l_k \alpha_l )\,,
\ee
where the coefficients $u_k^l$ are determined from the equation
\be
u^{n+1}_k -u_k^n u_k^1 -d_\pm^n (\nu ) u^{n-1}_k =0 \,.
\ee
This concludes the general scheme of expressing the potential term
in the action in
terms of the component fields $\phi_n$.

Using (\ref{bilpm}) and (\ref{bev})
one can easily  find the kinetic part of the action
\be
\label{kin}
S^\pm_{kin} = - \int d^2 z str\Big ( P^\pm \phi\partial_- \partial_+ \phi\Big )
= - \sum_{n=0}^\infty b^\pm_n \int d^2 z \phi_n^\pm \partial_- \partial_+
\phi_n^\pm \,, \ee
where
\be
\label{b} b_n^\pm =(-1)^n 2^{-(2n+1)}(n\,!)^2  \prod_{l=
-n}^n \left (1\mp \frac{2\nu}{2l+1} \right ) \,.  \ee

Let us now check these general formulae for the lowest finite systems.
For the particular cases of $2\nu =2l+1$,
where $l $ is an integer, the computation is simplified since
only a finite number of fields contribute because all terms involving
$P^\pm h_n$ with $n\geq l+\half \mp\half$ decouple from the supertrace.
As a particular manifestation of this phenomenon the coefficients (\ref{b})
all vanish starting from $n=l+\half\mp\half$. For $2\nu \neq 2l+1$ all
$b_n^\pm$ are non-vanishing. In this case however the coefficients with
$n$ high enough take both positive and negative values so that the full
infinite system involves ghosts.
Due to the invariance of the full original algebra with respect
to $\nu\lra -\nu$, $K\lra -K$ which for $\nu =2l+1$
is equivalent to $P^\pm \lra
P^\mp$, $l\lra -(l+1)$ we only need
to consider the case of non-negative $\nu$.

According to (\ref{gf1}) one needs to know the generating function
$Z^\pm (\nu ;\alpha )$ and the difference
\be
\label{fdif}
\hat{\phi}(z)=\dtp (z)-\phi (z)\,.
\ee
Let us first note that the components $\phi^\pm_0$ do not
contribute to the potential since $\dtp {}_0^\pm =\phi_0^\pm$, which
 is in fact obvious from (\ref{pot}). This is a particular manifestation
of the general property that if $\phi$ is such that $\phi_n^\pm =0$ for
$n\geq n_0$ then $\hat{\phi}_n^\pm =0$ for $ n\geq n_0 -1 $.

For $l=0$  one easily finds that the potential (\ref{pot}) $V^\pm (1;\phi )=0$
since the only non-vanishing component of $\hat{\phi}$ is
$\hat{\phi}_0 (z)$ and according to (\ref{mgf}) $Y^\pm_1 (0;\alpha )$=
$\frac{1}{2} (1\mp 1)$ so that inserting it back into (\ref{gf1})
one finds that the terms involving derivatives trivialize while
the constant factor turns out to be proportional to $(1\pm 1)$
and therefore the potential vanishes identically.

Let us so consider the case $l=1$, i.e. $2\nu =3$.
For the $``+"$ case one concludes from the formula (\ref{fact}) that
again the factor algebra is one-dimensional ($``A_0 "$),
i.e. only the constant
mode contributes under the supertrace leading to a vanishing
potential. Hence we now discuss the $``-"$ case of $A_1$ where the $h_1$
- dependent terms have to be taken into account.
{}From (\ref{gf}), (\ref{gfy}) and (\ref{mgf}) it follows that
\be
Z^- (3/2;\alpha )= e^{\alpha_0} Y^- (3/2;\alpha_1 )
\ee
with
\be
Y^- (3/2;\alpha ) =2 ch (\alpha )\,.
\ee
Inserting all this into (\ref{gf1}) and taking into account that
$\hat{\phi} {}^-_0 = 2\phi^-_1$ and $\hat{\phi} {}^-_1 =0$ one finds
\be
\label{pot1}
V^-(3/2 ;\phi^- )=2exp(2\phi^-_1 )\,.
\ee
With the proper rescaling of the $\phi_1 $ field we recognize the
Liouville potential.

Next we treat the case $l=2$, i.e. $2\nu =5$. The calculation in the
$``+"$ case of $A_1$ follows exactly the lines of the $``-"$
case for $l=1$. Also in
this case all $h_n$ with $n > 1$ decouple. The final result for the
potential only
differs in the overall sign. In the $``-"$ case also the $h_2$ terms
in $\phi^-$
couple in the supertrace. {}From (\ref{shiftT}) one easily finds
for arbitrary $\nu$
\be
\label{shift2}
T^- h_2 = (h_2 + 4h_1 +4 ) T^- \,.
\ee
{}From this equation and (\ref{shift0}) it follows
that
\be
\hat{\phi}^- =4\phi^-_2 h_1 +4\phi^-_2 +2\phi^-_1 \,.
\ee
The function $Y^\pm (\alpha )$ has the following form
\be
Y^- (5/2;\alpha )= \frac{sh(3\alpha )}{sh(\alpha )}=
e^{2\alpha} +1+e^{-2\alpha}\,.
\ee

{}From (\ref{gf1}) it now follows that
\be
\label{pot2}
V^- (5/2; \phi )
=4e^{2\phi^-_1}\left (e^{4\phi^-_2} +e^{-4\phi^-_2}\right )\,.
\ee

Finally we investigate the case $l=3$. The $``+"$ case
of $A_2$ is again
parallel to the $``-"$ case for $l=2$. As in that case $h_n$ with
$n>2$ decouple and the final expression for the potential only
differs from (\ref{pot2}) by a minus sign. In the $``-"$ case
also the $h_3$ terms couple.

The first step is to derive
$\hat{\phi}^-$. {}From (\ref{shiftT}) we find
\be
T^- h_3 =\left( h_3 +6 h_2 +12 h_1 +9+\frac{4\nu^2}{5}-
\frac{4\nu K}{5} \right )T^- \,.
\ee
Using also (\ref{shift2}) and (\ref{shif}) one finds
for $2\nu=7$ and  $K=-1$
\be
\label{ph3}
\hat{\phi^-}=  6\phi^-_3 h_2 +(12\phi^-_3 +4\phi^-_2 )h_1
+\frac{108}{5} \phi^-_3 +4\phi^-_2 +2\phi^-_1 \,.
\ee

The next step is to compute the generating function $Z^- (7/2;\alpha )$.
In order to do that we first rewrite $Y^- (7/2;\alpha)$ as
\be
Y^- (7/2; \alpha ) =e^{3\alpha}+e^\alpha +e^{-\alpha} +e^{-3\alpha}
\ee
so that
\be
Z^- (7/2;\alpha_0, \alpha_1 ,0,0,\ldots )=e^{\alpha_0}
\left (e^{3\alpha_1}+e^{\alpha_1} +e^{-\alpha_1} +e^{-3\alpha_1}\right )\,.
\ee
Now it is easy to integrate the equation (\ref{difa}) at $n=1$
to obtain the full generating function $Z^- (7/2; \alpha )$
\be
Z^- (7/2;\alpha )=e^{\alpha_0}\left (e^{4\alpha_2 +3\alpha_1} +e^{-4\ga_2
+\ga_1}
+e^{-4\ga_2 -\ga_1}+e^{4\ga_2 -3\ga_1} \right )\,.
\ee
Inserting this generating functional with $\hat{\phi}^-$ from
(\ref{ph3}) into (\ref{gf1}) gives the potential
\bee
\label{pot3}
V^- (7/2;\phi )&{}&
= 2 \Bigl( 3exp (\frac{48}{5} \phi^-_3 +8\phi^-_2 +2\phi^-_1 )\nn
&{}&+4exp (-\frac{72}{5} \phi^-_3 +2\phi^-_1 )+
3 exp (\frac{48}{5} \phi^-_3 -8\phi^-_2 +2\phi^-_1 ) \Bigr)\,.
\eee

In principle one can
derive
in this way
the potential for
any finite system. The feature that the potential for the $``+"$
case with $2\nu = 2l+1$ coincides with the negative of the potential
for the $``-"$ case with $2\nu =2l -1$ is
a general feature which follows from (\ref{fact}) and (\ref{mattr}).
In the sequel we only consider the $``-"$ cases.

We then compare
the various cases  to the standard Toda actions
\be
\label{actt}
S_l =-\int d^2 z \left (  \sum_{a,b =1}^l  K^{-1\,ab}q_a \partial_- \partial_+
q_b +\sum_{a =1}^l e^{-q_a } \right )\,,
\ee
where $K_{ab}$ is the $A_l$ Cartan matrix, i.e.
the non-vanishing components of $K_{ab}$ are $K_{aa} =2$, $K_{a,a-1
}=K_{a-1,a}=-1$.

For the Liouville case which corresponds to $l=1$ the action can always be
reduced to
the form (\ref{actt}).

For the case $l=2$ we find from  (\ref{pot2}) that the relevant
field redefinitions are
\be
q_1 =-2\phi_1^- -4\phi_2^- -ln 4\,,\qquad
q_2 =-2\phi_1^- +4\phi_2^- -ln 4\,.
\ee
With this substitution one arrives at the formula (\ref{actt}).

For the case $l=3$ we proceed in the same manner and find from
(\ref{pot3}) that the relevant field redefinitions are
\bee
q_1 &=&-\Big (\frac{48}{5} \phi^-_3 +8\phi^-_2 +2\phi^-_1 +ln6 \Big )\,,\nn
q_2 &=&-\Big (-\frac{72}{5} \phi^-_3 +2\phi^-_1 +ln8 \Big )\,,\nn
q_3 &=&-\Big (\frac{48}{5} \phi^-_3 -8\phi^-_2 +2\phi^-_1 +ln6 \Big )\,.
\eee
By substituting these field variables into the action we again find the
canonical
form (\ref{actt}).

Let us also mention that one can consider the one-dimensional Toda systems
within
this formalism too. The equation of motion then reads
\begin{equation}
\ddot{\phi} = [H^+ ,H^- ]\,.
\end{equation}

By introducing the canonical momentum $\pi$ one arrives at the Hamiltonian
equations
\begin{equation}
\dot{\phi} = \pi\,, \qquad \dot{\pi}  = [H^+ ,H^- ]
\end{equation}
and it is easy to formulate it in a Lax pair form
\begin{equation}
L = \pi I + \frac{1}{\sqrt{2}}K (H^+ + H^- )\,,
\end{equation}
\begin{equation}
A = \frac{1}{\sqrt{2}}K (H^+ -H^- )\,,
\end{equation}
\begin{equation}
\dot{L} = [A, L]\,.
\end{equation}

\setcounter{equation}{0}

\section{Higher-spin symmetries}

The generalized Toda field equations were derived in (\ref{equ}) from the
condition that the
higher-spin curvatures (\ref{R}) be zero. It is then natural to seek the
symmetries of the Toda equations as specific higher spin gauge transformations
\begin{equation}
\label{hsg}
\delta W_\mu = \partial_\mu \epsilon +[W_\mu , \epsilon ]\,.
\end{equation}
For example the conformal transformations
of the $\phi$ - field (\ref{conf})
can be derived
from the gauge transformations
as follows.
Let us consider gauge parameters of the form
\begin{equation}
\label{dec}
\epsilon = \epsilon^+ + \epsilon^0
\end{equation}
with
\begin{equation}
\epsilon^+ =u(z)T^+ \,,\qquad \epsilon^0 =v(z) T^0\,.
\end{equation}
One can then ask  which gauge transformations
leave invariant the Toda ansatz (\ref{ans}) and (\ref{gauge}) for gauge fields.
Note that the gauge condition (\ref{gauge}) is a convenient one because it
implies that one has to require
\begin{equation}
\label{cond} \delta
W_\mu^{positive} =0\,,
\end{equation}
where $W_\mu^{positive}$ denotes all gauge
fields with positive gradings with respect to $T^0$.  (The gauge condition
(\ref{gauge}) tells us that the field $H^+$ is constant and therefore its
variation vanishes.) When analyzing the anti-conformal invariances it is more
convenient to fix another gauge $\phi_- =0$.

The only non-trivial condition for the simplest case under
consideration is
\begin{equation}
\label{ch}
0=\delta W_\mu^+ =\partial_\mu \epsilon^+ +[W_\mu^+ ,\epsilon^0 ]
+[W^0_\mu , \epsilon^+ ]\,,
\end{equation}
where $W_\mu^+ =\delta_\mu^+ H^+$ and $W_\mu^0 =\omega_\mu (T^0 )$.
Now setting $\mu =-$ and taking into account that
$W_-^0 =0$ due to the
gauge condition (\ref{gauge}) one finds
\begin{equation}
\partial_- \epsilon^+ =0\,.
\end{equation}
For $\mu =+$ one finds taking into account that $H^+ = T^+$
due to (\ref{hpm}) and the gauge condition (\ref{gauge}):
\begin{equation}
0=\partial_+ \epsilon^+ + [T^+ , \epsilon^0 ] +[\partial_+ \phi ,\epsilon^+
]\,.
\end{equation}
One can solve this for $\epsilon^0$ with the result
\begin{equation}
\epsilon^0 =\partial_+ \phi u(z^+ )+T_0 \partial_+ u(z^+ )
\end{equation}
that gives exactly the conformal transformation (\ref{conf}) for $\phi$,
taking into account that
\begin{equation}
\delta H^- = [H^- ,\epsilon^0 ]
\end{equation}
because there are no negative terms in $\epsilon$.

The above analysis generalizes to
higher-spin conformal transformations as follows.
The decomposition (\ref{dec}) relevant for the conformal
symmetry generalizes to
\begin{equation}
\label{decomp}
\epsilon^s =\sum_{n=0}^{s-1} \epsilon^{s,n}\,,
\end{equation}
where
\begin{equation}
[T^0 ,\epsilon^{s,n} ]=n \epsilon^{s,n}
\end{equation}
and $s$ is the spin of the corresponding higher-spin current
so that $s=2$ for the case of conformal symmetry.

As before we fix the gauge (\ref{gauge}) and impose the conditions
(\ref{cond}).
This leads to the following chain of equations generalizing (\ref{ch}):
\begin{equation}
0=\delta W_\mu^n =\partial_\mu \epsilon^{s,n} +
[W_\mu^+ ,\epsilon^{s,n-1} ] +
[W_\mu^- ,\epsilon^{s,n+1} ] +
[W^0_\mu , \epsilon^{s,n} ]
\end{equation}
for $n>0\,,\, \mu = +$ and $n\geq 0\,,\,\mu = -$. As a result one finds,
respectively
\begin{equation}
\label{basic}
0=\delta W_+^n =\partial_+ \epsilon^{s,n} +
[T^+ ,\epsilon^{s,n-1} ] +
[\partial_+ \phi , \epsilon^{s,n} ]\,;\qquad n>0
\end{equation}
and
 \begin{equation}
\label{second}
0=\delta W_-^n =\partial_- \epsilon^{s,n} +
[H^- ,\epsilon^{s,n+1} ]\,;\qquad n\geq 0
\end{equation}
taking into account (\ref{connec}) for the particular gauge (\ref{gauge}).

One observes that for the terms with highest grading in the expansion
(\ref{decomp}) the consistency of the above equations requires,
respectively
\begin{equation}
[T^+ , \epsilon^{s,s-1} ]=0
\end{equation}
and
\begin{equation}
\label{global}
\partial_- \epsilon^{s,s-1} =0\,,
\end{equation}
i.e. the top parameters $\epsilon^{s,s-1}$  must be $z^-$ - independent
highest weight vectors of the basic $sl_2$ algebra. $\epsilon^{s,s-1}(z^+ )$
serve as independent parameters of the higher-spin symmetries.

One can then use the basic relations (\ref{basic}) to reconstruct
all lower $\epsilon^{s,n}$ with $0<n<s-1$ in terms of $\epsilon^{s,s-1}$
and ($\partial_+$ - derivatives of) $\phi$,
\begin{equation}
\label{recur}
\epsilon^{s,p} = -t({\cal D}\epsilon^{s,p+1}) +   \epsilon^{s,p}_+ \,,
\end{equation}
where $\epsilon^{s,p}_+ $ are some  quantities obeying the
condition
\begin{equation}
\label{hw}
[T^+ , \epsilon^{s,p}_+ ] =0
\end{equation}
(i.e. highest weight vectors of $sl_2$),
${\cal D}$ is the Lorentz covariant derivative
\begin{equation}
\label{cov}
{\cal D} A=\partial_+ A + [\partial_+ \phi , A]
\end{equation}
and
$t$ is defined in such a way that
\begin{equation}
\label{inver}
[T^+ , t(A)] = A
\end{equation}
for any operator $A\in A(2;2\nu )$ which has a non-negative grading (and not a
constant in the Weyl ordering).

In what follows we use the following definition for $t$ in the basis
of Weyl ordered polynomials used in (\ref{base})
\begin{equation}
\label{t}
t(E_{n,m})=-{1\over m+1}E_{n-1,m+1}\quad  n>0\,;\qquad  t(E_{0,m})=0\,.
\end{equation}
An important property of this operation is that it is skewsymmetric
in the following sense
\be
\label{skew}
 str(At(B))=- str(t(A) B)\,.
\ee

It is easy to see that such a definition of $t$ leads to the following
 relations
\begin{equation}
\label{comt}
[T^+ ,t(x)] =x-\Pi_- (x) \,,\qquad
t([T^+ ,x]) =x-\Pi_+ (x)\,,
\end{equation}
where $\Pi_+$ and $\Pi_-$ are respectively projection operators to
highest weight and lowest weight components, i.e.
\begin{equation}
\label{project-}
\Pi_- (E_{0,m})=E_{0,m}\,;\qquad
\Pi_- (E_{n,m})=0\quad n>0\,,
\end{equation}
\begin{equation}
\label{project+}
\Pi_+ (E_{n,0})=E_{n,0}\,;\qquad
\Pi_+ (E_{n,m})=0\quad m>0\,.
\end{equation}

Also in what follows we will use the following
projection operator on the neutral subspace
\begin{equation}
\label{project0}
\Pi_0 (E_{n,n})=E_{n,n}\equiv h_n\,;\qquad
\Pi_0 (E_{n,m})=0\quad n\ne m\,.
\end{equation}

The iterative solution of equation (\ref{recur}) (equivalently (\ref{basic}))
for $\epsilon^{s,0}$ can be written in the following form
\begin{equation}
\label{iter}
\epsilon^{s,0}=\sum_{p=0}^{s-1} (-t\D )^p \epsilon_+^{s,p}\,,
\end{equation}
where $\epsilon_+^{s,s-1}$=$\epsilon^{s,s-1}$.

In order to guarantee that (\ref{second}) is also true it suffices
to ensure that
\begin{equation}
\label{top}
\Pi_+ (\partial_- \epsilon^{s,p} +[H^- ,\epsilon^{s,p} ])=0\,,
\end{equation}
and take into account the Toda equations. Indeed, using
the fact that the
equations of motion result from the zero-curvature
conditions one observes that
\begin{equation}
[\partial_- +H^- ,
[{\cal D} + T^+ ,
\epsilon ]]
- [{\cal D} + T^+ ,
[\partial_- +H^- ,
\epsilon ]]
 =0\,,
\end{equation}
when (\ref{eq}) is true. Taking into account (\ref{basic}) one finds
that
\begin{equation}
\label{*}
[T^+ , (\partial_- (\epsilon )+[H^- ,\epsilon ])]=- {\cal D}(\partial_-
(\epsilon )+[H^- ,\epsilon ])\,.
\end{equation}
{}From here it follows that the components of $ (\partial_- (\epsilon )+[H^-
,\epsilon ])$ with all possible gradings vanish provided that the highest
weight components do (that allows one to start the inductive proof with the
vanishing right-hand side of (\ref{*}) with the highest possible grading).
This is indeed guaranteed by the condition (\ref{top}).

In order to satisfy (\ref{top}) one has to fix the free parameters
$\epsilon_+^{s,p}$ in (\ref{iter}) in an appropriate way. In
the next section we will give a constructive method which will
fix these parameters.

Since $\epsilon$ in (\ref{decomp}) is supposed to involve only
non-negative gradings, one finds from the transformation law of
$H^-$ that
\begin{equation}
\label{^}
\delta H^- = [H^- , \epsilon^{s,0} ]
\end{equation}
and therefore for a spin $s$ higher-spin transformation
\begin{equation}
\delta \phi = \epsilon^{s,0}\,,
\end{equation}
i.e. the parameters $\epsilon ^{s,0}$ serve as spin $s$ variations
of the physical fields.

So far we have studied the invariance at the level of the equations
of motion.
Let us investigate whether this invariance
also works for the action (\ref{action}). It will turn out that
the potential and kinetic terms in the action (\ref{action}) are invariant
separately. Indeed, the variation of the potential term is
\begin{equation}
\label{<}
\delta^{pot} S^\pm= 2 str\int d^2 z (P^\pm H^+ [H^- , \epsilon^{s,0} ] )\,,
\end{equation}
where we have taken into account that in the gauge under consideration
$H^+ =T^+$ and the variation of $H^-$ is given in (\ref{^}).
With the aid of the basic property of supertrace (for even higher spin
parameters) one transforms (\ref{<}) to
\begin{equation}
\delta^{pot} S^\pm = -2 str\int d^2 z (P^\pm [H^+ , \epsilon^{s,0} ] H^- )
\end{equation}
and then by virtue of the basic relations
(\ref{basic}) to
\bee
\delta^{pot} S^\pm =
2 str\int d^2 z (P^\pm {\cal D}( \epsilon^{s,1}) H^- )
=-2 str\int d^2 z (P^\pm \epsilon^{s,1} {\cal D}(H^- ))
\eee
(recall that $ {\cal D}(H^- )=R^-_{+-}=0$ is the defining relation for the
Lorentz connection $\omega_+$).

Now let us consider the kinetic term. According to (\ref{action})
\begin{equation}
\delta       S^\pm_{kin} = -2 \int d^2 z  str
(P^\pm \epsilon^{s,0} \partial_+ \partial_- \phi )\,.
\end{equation}
This variation would reduce to an integral of a full $\partial_-$
derivative if there exist such higher-spin currents $J^s (z^- )$, the
functionals of $\phi (z^+ ,z^- )$ at fixed $z_-$ and of higher spin symmetry
parameters, that their variation with respect to the dynamical fields
$\phi$ has the form
\begin{equation}
\label{curv}
\delta J^{s\pm} (z^- ) =\int dz_+ str (P^\pm \epsilon^{s,0} \partial_+
\delta\phi )\,.
\end{equation}
Indeed, in this case the variation of the kinetic term amounts to
the total derivative
\begin{equation}
\delta^{kin} S^\pm =-2 \int dz^- \partial_- J^{s\pm}\,.
\end{equation}

The higher-spin currents (\ref{curv}) indeed exist
analogously to the case of the ordinary Toda lattice
\cite{int2}-\cite{conf3},
\cite{scur,exp}.
The  relevant language here is that of  Hamiltonian dynamics.
Namely, we identify $z^-$ with the time coordinate while $z^+$ is
to be regarded as a spatial one. Later on we will assume that
all relations are valid for equal time variable $z^-$ which is not
written down explicitly in most cases. Also, we will sometimes use
$z$ instead of $z^+$. In accordance with (\ref{action}) the explicit
form of the Hamiltonian is
\begin{equation}
\label{ham}
{\cal H}^\pm =-2\int dz str (P^\pm H^+ H^- )\,.
\end{equation}

Let us introduce the notation
\begin{equation}
\label{dual}
\langle\xi , \phi \rangle =\int dz str \xi (z) \phi (z)\,,
\end{equation}
where $\xi (z)$ is an arbitrary grade zero element of the algebra $Aq(2;2\nu )$
regarded as a field independent parameter and having sufficiently smooth
behavior at infinity in $z$ so that one can freely integrate by parts any
terms involving $\xi$. Note that according to the defining property of
the supertrace
(\ref{strprop}) one has for the integer spin case under consideration that
\begin{equation}
\langle\xi , \eta \rangle = \langle \eta ,\xi \rangle\,.
\end{equation}
Also let us use the notation
\begin{equation}
\phi^\prime = \partial_+ \phi\,.
\end{equation}
So that, according to (\ref{dual})
\begin{equation}
\langle\xi^\prime ,\eta \rangle +
\langle\xi ,\eta^\prime \rangle =0\,.
\end{equation}

Next we define Poisson brackets $\{ ,\}$ for the dynamical variables $\phi$
as
\begin{equation}
\label{brack}
\{\langle\xi , \phi \rangle ,\langle\eta , \phi^\prime \rangle\}=
\langle\xi ,\eta \rangle
\end{equation}
for field independent $\xi$ and $\eta$.

Since the right hand side of (\ref{brack}) is field independent
the Jacobi identities are trivially satisfied.
It is worth mentioning that when the bilinear form (\ref{dual})
is nondegenerate, the definition (\ref{brack}) induces a Poisson
bracket structure on the dynamical variables $\phi$ themselves.
In particular, for this case
\begin{equation}
\label{shift}
\{\langle\xi , \phi \rangle , \phi^\prime \}=\xi\,.
\end{equation}
When the bilinear form (\ref{dual}) degenerates, the relation
(\ref{brack}) induces a Poisson bracket structure on the factor
algebra over null vectors of the bilinear form (\ref{dual}). This is
enough because the ideals formed by null vectors of the
bilinear form (\ref{dual}) do not contribute to any expression involving the
supertrace and in particular to the action (\ref{action}). In other
words one can always use (\ref{shift}) inside the supertrace operation.

Our aim is to prove that there exist such currents that
\begin{equation}
\label{curham}
\{\langle\xi , \phi \rangle , J^s \}=\langle\xi ,\epsilon^{s,0} \rangle\,,
\end{equation}
since this condition is equivalent to the condition (\ref{curv}),
i.e. the vector fields $\epsilon^{s,0}$, which act in the space of
the field variables $\phi$, are Hamiltonian. In the sequel we will not
explicitly indicate the $\pm$ projections by $P^\pm$. All calculations will
be insensitive to these projectors and they can always be introduced in the
final formulae.

The consistency conditions for this equation require
\begin{equation}
\label{Cons}
\{ \langle\eta ,\phi \rangle,\langle\xi ,\epsilon^{s,0} \rangle\}=
\{ \langle\xi ,\phi \rangle,\langle\eta ,\epsilon^{s,0} \rangle\}\,.
\end{equation}
Provided that the consistency conditions are satisfied,
one can derive the following explicit formula for $J^s$ in terms
of $\epsilon^{s,0}$
\begin{equation}
\label{cur1}
 J^s =\int_0^1 d\tau \langle\phi^\prime ,\epsilon^{s,0}(\tau\phi )
\rangle\,.
\end{equation}
Indeed, inserting this into (\ref{curham}) and using (\ref{Cons})
one easily shows that the left hand side of (\ref{curham}) amounts to
\begin{equation}
\int_0^1 d\tau {\partial\over\partial\tau}(\tau\langle \xi
,\epsilon^{s,0}(\tau\phi ) \rangle) =\langle\xi ,\epsilon^{s,0}
\rangle\,.
\end{equation}

In the next section we will describe a general scheme which will
allow us to prove the existence of such currents and at the same time
will provide a method to construct these currents. The fact that
the generalized Toda system under investigation admits a conformal
system of conserved higher-spin currents is not surprising of course
since the analogous property is well-known for the ordinary Toda lattice
field theory \cite{conf1}-\cite{conf3}. However an advantage of the method
described in the next section is that it can be practically useful for the
derivation of the explicit form of particular currents.

\setcounter{equation}{0}

\section{Governing Equation}

Let us now introduce the equation that governs the structure
of all the higher spin currents. In order to do that it is
convenient to introduce generalized currents
$J^s (\phi ,\mu )$ which depend on
a new parameter $\mu \in Aq (2;2\nu )$ such that $\Pi_- \mu$=$\mu$, i.e. $\mu$
is an arbitrary lowest weight vector of the $sl_2$ algebra of the form
\be
\label{mu}
\mu =\sum_{n=-\infty}^\infty \sum_{m=0}^\infty z^n (T^- )^m \mu_{n,m}\,,
\ee
where $\mu_{n,m}$ are arbitrary coefficients.
The $\mu$ - independent part of the generalized currents
$J^s (\phi ,0)$ will
be identified with the higher spin currents in (\ref{curham}).
The governing equation has the form
\begin{equation}
\label{gov}
\Dc_\xi J^s =0\,,
\end{equation}
where
\begin{equation}
\label{cd}
\Dc_\xi =
\langle\xi ,
\frac{\delta}{\delta\phi^\prime} \rangle
- \langle \Pi_- U \xi ,
\frac{\delta}{\delta\mu}
 \rangle\,,
\end{equation}
 $\xi$ is an arbitrary grade zero element of $Aq(2;2\nu )$,
\begin{equation}
\label{U}
U=(1- D t)^{-1}\,,
\end{equation}
\begin{equation}\label{der}
D(X)=\partial_z (X)+ [\phi^\prime ,X] +[\mu ,X]
\end{equation}
and the variational derivatives are defined in the natural way as
\be
\label{vari}
\langle \xi ,
\frac{\delta}{\delta\phi^\prime}\rangle \phi^\prime  =\Pi_0 (\xi )\,;\qquad
\langle \eta ,
\frac{\delta}{\delta\mu}\rangle \mu  =\Pi_- (\eta ) \,.
\ee
For future use we also introduce
\begin{equation}
\label{V}
V=(1- tD )^{-1}\,,
\end{equation}
\begin{equation}
\label{G}
G=tU=Vt\,.
\end{equation}
The latter equation can be easily derived by expanding (\ref{U}) and
(\ref{V}) in power series, as well as
\begin{equation}
\label{U-1}
U-1=\,D \,G
\end{equation}
and
\begin{equation}
\label{V-1}
V-1=\, GD\,.
\end{equation}
Another useful formula is
\be
\label{usef}
\delta (UD) =
\delta (DV) = U\delta (D) V\,,
\ee
where we have taken into account that $t$ is field independent.

As shown below,
the variational equations (\ref{gov}) are indeed
consistent in the following standard sense:
\be
\label{cons}
\Dc_{\xi_1}\Dc_{\xi_2}=
\Dc_{\xi_2}\Dc_{\xi_1}\,.
\ee

If we rewrite (\ref{gov}) symbolically as
\begin{equation}
\label{c}
\xi^i{\partial\over \partial x^i } J =\xi^i A_i (J)\,,
\end{equation}
where $x_i$ stands for $\phi^\prime$ while
\be
\label{A}
\xi^i A_i =
 \langle \Pi_- U \xi ,
\frac{\delta}{\delta\mu}\rangle
\ee
 is the second term of the linear operator
on the left hand side of (\ref{gov}), then equation (\ref{cons}) is
equivalent to the following zero-curvature conditions
\begin{equation}
\label{Cur}
{\partial\over \partial x^i} A_j
-{\partial\over \partial x^j} A_i =[A_i ,A_j ]\,.
\end{equation}

By using (\ref{A}) the explicit form of this condition is
\begin{equation}
\label{start}
Y=
\langle\xi_1 ,
\frac{\delta}{\delta\phi^\prime } \rangle \eta_2
-\langle
\eta_1, \frac{\delta}{\delta\mu}\rangle
\eta_2
-(1\lr 2)
=0\,,
\end{equation}
where
\begin{equation}
\label{eta}
\eta_i =\Pi_- U \xi_i\qquad i=1,2\,.
\end{equation}
Let us now prove that $Y=0$.
Using the explicit form of the operator
$ U$
one transforms
$Y$  to the following form
\begin{equation}
\label{Y1}
Y= \Pi_- U\Big ([\xi_1 ,G\xi_2 ] -
[\Pi_- U\xi_1 ,G\xi_2 ]\Big ) -(1\lr 2)\,.
\end{equation}
Now we use the identity (\ref{Uid}) proven in Appendix A
for the first term in the above equation
to arrive at the following expression
\bee
\label{Y2}
Y&=&
\Bigl(  \Pi_-\Big ([U\xi_1 ,G\xi_2 ] +
D G([\Pi_- U\xi_1 ,G\xi_2 ])\Big )\nonumber\\
&-&\Pi_- U
[\Pi_-U\xi_1 ,G\xi_2 ] \Bigr)-(1\lr 2)\,,
\eee
where the last term in (\ref{Uid}) trivializes since $\xi_i$ is grade zero.
The crucial point is then that the first term in the above equation
can be rewritten as
\bee
\label{Y4}
 \Pi_-
[U\xi_1 ,tU\xi_2 ] &-&(1\lr 2)
=\Big (\half
 \Pi_- [T^+ ,
[tU\xi_1 ,tU\xi_2 ]]\nonumber\\
&+&
 \Pi_-
[\Pi_- U\xi_1 ,tU\xi_2 ]\Big ) -(1\lr 2)\,.
\eee
The first term in this expression is zero since $\Pi_-([T^+ ,X])=0$
for any $X$ due to (\ref{project-}).
Inserting the remaining term in (\ref{Y4})
into (\ref{Y2}) and taking into account (\ref{U-1})
we find that the terms cancel
pairwise. We have thus proven that the governing equation (\ref{gov})
is indeed consistent.

{}From the general representation (\ref{Cur})
one can write down a general solution
for $J$ in the form
\begin{equation}
\label{gens}
J=P\exp [\int_0^1 ds x^i A_i (sx )]\,\,j\,,
\end{equation}
where $j$ is an arbitrary $x $- independent quantity and the $P$
ordering is with respect to $s$.

In order to implement this formula for the $A_i$ of (\ref{A}) one has just
to replace the parameter $\xi$ in (\ref{A}) by $\phi^\prime$.
This leads to the result
\begin{equation}
\label{sol}
J (\phi ,\mu )=P\exp [\int_0^1 ds  \langle \Pi_- U (s\phi^\prime )
(\phi^\prime ) ,
\frac{\delta}{\delta \mu}\rangle ]
\,\,j (\mu )\,,
\end{equation}
where $j(\mu )$ is an arbitrary complex valued functional of $\mu$.

The main properties of this current which we will prove in
Appendix B can be summarized as follows:

(i) The $\mu$ -independent part $J(\phi ,0)$ of the current (\ref{sol}) is
conserved according to the Hamiltonian
     equations of the Toda system for any $\phi$ - independent
     and time independent functional $j(\mu )$. The functional $j(\mu )$
parametrizes the full space of conserved currents. In what follows we
refer to the functional $j(\mu )$ as the $j$-parameter.

(ii) {}From (\ref{sol}) it follows that
\begin{equation}
\label{phi0}
j(\mu )=J(0,\mu )\,.
\end{equation}

(iii) Since the governing equation (\ref{gov})) is a first-order
variational equation,
according to the standard Leibniz rule, a product $J_{1\times 2}=J_1 J_2$
satisfies the
governing equation if each of $J_1$ and $J_2$ do. {}From (\ref{phi0})
it follows then that
\begin{equation}
\label{1x2}
j_{1\times 2}(\mu )=j_{1}(\mu )j_{2}(\mu )\,.
\end{equation}
This extends to any function $f(J)$ which can be
represented as a power series. The $j$-parameter is
then represented by $f(j)$.

(iv) Given two currents $J_1$ and $J_2$ satisfying the governing
equation (\ref{gov}), the bilinear combination
\begin{equation}
\label{1u2}
J_{1, 2}(\phi ,\mu )=\langle \frac{\delta J_{1}}{\delta \mu},
UD(\frac{\delta J_{2}}{\delta \mu})\rangle
\end{equation}
satisfies the governing equation (\ref{gov}) too.

(v)  The Poisson bracket for the two conserved currents with respect
    the standard Poisson structure of $\phi$ variables (\ref{brack})
is indeed a current of the form $J_{1,2}$, i.e.
\begin{equation}
\label{1,2}
\{J_1 ,J_2 \}_\phi \vert_{\mu =0} =J_{1,2}\vert_{\mu =0}\,.
\end{equation}
According to (iv) this shows that, as expected, the
Poisson bracket of two conserved currents gives a new conserved current
and $J_{1,2}$ gives an appropriate solution of the governing equation
that corresponds to this conserved current.

(vi) According to (\ref{1u2}) the $j$-parameter for the conserved current
$\{J_1 ,J_2 \}_\phi \vert_{\mu =0}$ therefore is
\begin{equation}
\label{j12}
j_{1, 2}(\mu )=\langle \frac{\delta j_{1}}{\delta \mu},
U_0  D_0 (\frac{\delta j_{2}}{\delta \mu})\rangle
\end{equation}
with
\begin{equation}
\label{U0}
U_0 =(1- D_0 t)^{-1}\,,\qquad
V_0 =(1- t D_0 )^{-1}\,,
\end{equation}
\begin{equation}
\label{der0}
D_0 (A)=\partial_z (A) +[\mu ,A] \,.
\end{equation}
The right hand side of (\ref{j12}) defines a Poisson
structure in the space of $\mu$ variables
\begin{equation}
\label{mbrac}
\{j_1 (\mu ),j_2 (\mu )\}_\mu =\langle \frac{\delta j_{1}}{\delta \mu },
U_0  D_0 (\frac{\delta j_{2}}{\delta \mu})\rangle\,.
\end{equation}

The fact that the Jacobi identities are satisfied for this
Poisson structure follows directly from the uniqueness
of the solution of the governing equation and the fact that
the original $\{,\}_\phi$ bracket respects the Jacobi
identities. Nevertheless we will give an independent proof
of the Jacobi identities for (\ref{mbrac}) in Appendix B.

The propositions (ii), and (iii) are
straightforward and follow from the arguments of
the text. The proofs of the propositions (i), (iv)-(vi)
are given in Appendix B.

Note that the description of the currents in terms of a lowest
weight vector field $\mu (z)$ is a different representation
as compared to the zero-grade field $\phi (z)$.
It is the Poisson brackets (\ref{mbrac}) between the $j$-parameters
from which
one can directly read off the algebra of conserved currents
in the Toda system.
Since the Poisson structure (\ref{mbrac}) is
non-linear in $\mu$, from (\ref{1,2}) it follows that this
non-linearity just reproduces a non-linear structure of the
form of the Poisson algebra of conserved currents.
The two representations are dual to each other in a certain sense.
In the $\phi$ - representation the Poisson structure is simple but
the form of the currents is quite complex.
In the $\mu$ - representation the Poisson structure on the other hand
is nonlinear but the conserved currents are very simple.
We will illustrate this in specific examples later.

Let us now discuss the relation between the generalized current
(\ref{sol}) and the higher spin currents discussed in the
previous section.
{}From the property (iii) it follows that
in order to construct a complete set of
conserved currents it is enough to consider $j$ - parameters
linear in $\mu$.

We first note that the current (\ref{sol}) satisfies the following
formula
\bee
\label{f}
J (\phi ,\mu )
&=&
J (0 ,\mu ) +
\int_0^1 ds  \langle \Pi_- U (s\phi^\prime )
(\phi^\prime ) , \frac{\delta}{\delta \mu}\rangle
P\exp [\int_0^s ds^\prime  \langle \Pi_- U (s^\prime\phi^\prime )
(\phi^\prime ) , \frac{\delta}{\delta \mu}\rangle ]
\,\,j (\mu )\nn
&=& j (\mu) +
\int_0^1 ds  \langle \Pi_- U (s\phi^\prime )
(\phi^\prime ) , \frac{\delta}{\delta \mu}\rangle
J (s\phi ,\mu )\,.
\eee
The first equality in (\ref{f}) can be obtained from (\ref{sol})
by using a power series expansion for a path ordered exponential.
The second equality follows from a rescaling of the integration
parameter $s^\prime$. The physical currents are identified with
the (conserved) $\mu$-independent part of
$J (\phi ,\mu )$. {}From (\ref{f}) it follows that
\be
\label{ff}
J(\phi ,0)=
 j (0) +
\int_0^1 ds  \langle \Pi_- {\cal U} (s\phi^\prime )
(\phi^\prime ) , \epsilon_+ (s\phi^\prime )\rangle\,,
\ee
where
\be
\label{fff}
\epsilon_+ (\phi )=
\frac{\delta}{\delta \mu}
J (\phi ,\mu )\vert_{\mu=0}
\ee
and
\be
\label{cU}
{\cal U}= (1-\D t )^{-1}\,,\qquad \D =\partial_z +[\phi^\prime ,
\phantom{M}]
\,.
\ee
Note that $\epsilon_+$ is a highest weight vector since
$\mu$ is a lowest weight vector.
The effect of $\Pi_-$ in (\ref{ff}) is unity because
$\Pi_+ (\epsilon_+ )$=
$\epsilon_+$. Furthermore the operator ${\cal U}$ can be partially integrated
so that one is left with
\be
\label{ffff}
J(\phi ,0)=
 j (0) +
\int_0^1 ds  \langle
\phi^\prime  ,
{\cal V} (s\phi^\prime )
\epsilon_+ (s\phi^\prime )\rangle\,,
\ee
where
\be
\label{cV}
{\cal V}= (1-t \D  )^{-1}\,.
\ee
This equation is in fact
the equation (\ref{cur1}) with the solution for
\be
\label{e0}
\epsilon^{s,0}=\Pi_0 \left (
{\cal V} (\phi^\prime )
\epsilon_+ (\phi^\prime ) \right )
\ee
from
the equation (\ref{iter}). Note that the constant $j(0)$ which appears in
(\ref{ffff}) does not affect the transformation properties. As is shown below
this constant is related to the classical central charge in the model.

\setcounter{equation}{0}

\section{Some Particular Currents from the
 Governing Equation}

Let us illustrate how one can apply the governing equation
in practical calculations.
As a simplest example
we consider the
case of the spin -  1 current.
In this case the $j$- parameter has the form
\be
\label{j1}
j (\mu )=
\langle v  ,\mu\rangle\,,
\ee
where the $c$-number parameter $v=v(z)$ is arbitrary. Note that only the zero
grading part of $\mu$ is important.  Inserting this formula into (\ref{sol})
and
considering the $\mu$ - independent part we arrive at
\be
\label{JJ1}
J^1(\phi ,0) =
\int_0^1 ds
\langle \Pi_- {\cal U} (s\phi^\prime )
(\phi^\prime ) ,
\frac{\delta}{\delta \mu}\rangle\,
\langle v  ,\mu\rangle\vert_{\mu =0}=
\int_0^1 ds
\langle v  ,{\cal U} (s\phi^\prime ) (\phi^\prime )
\rangle =\int dz v(z) str(\phi^\prime)\,.
\ee
This current is indeed trivial since the mode $str(\phi^\prime )$
decouples from the other fields in all expressions. It hence
must have trivial Poisson brackets to all other currents of the theory.

Next we apply the formulae for the case of the conformal transformations.
In this case the $j$- parameter has the form
\be
\label{j2}
j (\mu )=
\langle u T^+ ,\mu\rangle\,,
\ee
where $u=u(z)$ is an arbitrary conformal parameter and $\mu=\mu (z)$
is an arbitrary grade $-1$ lowest weight variable. (Let us remind ourselves
that the definition (\ref{dual}) involves an integration over $z$.)
Inserting this formula into (\ref{sol}) and considering the $\mu$ - independent
part we arrive at
\be
\label{JJ}
J^2(\phi ,0) =
\int_0^1 ds
\langle \Pi_- {\cal U} (s\phi^\prime )
(\phi^\prime ) ,
\frac{\delta}{\delta \mu}\rangle\,
\langle u T^+ ,\mu\rangle\vert_{\mu =0}=
\int_0^1 ds
\langle u T^+ ,{\cal U} (s\phi^\prime ) (\phi^\prime )
\rangle\,.
\ee
We notice that we only need to consider the linear term in the exponential
because all the higher order terms will necessarily contain higher negative
gradings which cannot contribute under the supertrace in the case under
consideration where $T^+$ is the only operator with positive grading.
Similarly we only need to consider the second term in the expansion of
${\cal U}$. A straightforward calculations then leads to the following
expression
for the current
\be
\label{JJJ}
J^2 (\phi ,0)=-\int dz str (\half u(z) (\phi^\prime )^2 +u^\prime (z) T^0
\phi^\prime )
\,,
\ee
which we recognize as the current leading to the transformation (\ref{conf}).

The natural conjecture for the spin - $s$ current is
\be
\label{js}
j_v^s (\mu )=
\langle v (T^+ )^{s-1} ,\mu\rangle\,,
\ee
where $v=v(z)$ is an arbitrary parameter.
Analogously one can use the formula (\ref{sol}) along with the property
(i) for the derivation of the explicit form of the higher-spin currents.
We hope to describe some more examples elsewhere. Here let us just note
that from (\ref{sol}) it follows that only a finite number of terms is
contributing for any fixed spin. The reason is that each term in the
power series expansion of (\ref{sol}) contains the negatively graded
operator $t$ so that only a finite number of terms
can match a limited positive grade in the original spin-$s$ $j$-parameter
(\ref{js}).

As an alternative to checking the algebra of the currents we can equally
well do it for the $j$ - parameters using (\ref{mbrac}). As an example,
let us check the conformal algebra in these terms.
\begin{equation}
\label{mbrac1}
\{j_1 (\mu ),j_2 (\mu )\}_\mu =
\langle
u_1 T^+,
U_0  D_0 (u_2 T^+ )\rangle =
\langle
u_1 T^+,
(D_0 t +D_0 tD_0 t )
(u_2^\prime  T^+ )
+(1+D_0 t)u_2 [{\mu}, T_+ ]
\rangle\,,
\end{equation}
where again only the first few terms in the expansion of $U_0$
contribute. After some algebra one finds
\be
\label{s2}
\{j_1 (\mu ),
j_2 (\mu )\}_\mu =
\int dz (u_1^\prime u_2 -u_2^\prime u_1 )str\left (T^+ \mu (z)\right )
+\half \int dz u_1 u_2^{\prime\prime\prime} str(T^+ T^- )\,.
\ee
Reminding ourselves that we can project out with $P^\pm $ we can compute the
final supertrace to obtain
\be
\label{ss2}
\{j_1 (\mu ),j_2 (\mu )\}_\mu =
j_{1,2}
+\frac{1}{32} (1-4\nu^2 )\left (1\mp \frac{2\nu}{3} \right )
 \int dz u_1 u_2^{\prime\prime\prime}
\ee
with the convention that the upper (lower) sign corresponds to
the $P^+$ ($P^-$) sector and $j_{1,2}$ is the spin two current
with the parameter $u_{1,2}=u_1^\prime u_2 -u_2^\prime u_1$.

Let us now verify that the spin-$s$ current (\ref{js})
indeed transforms as a
primary spin - $s$ field
under the conformal transformations.
\begin{equation}
\label{mbracs}
\{j_u^2 (\mu ),j_v^s (\mu )\}_\mu =
\langle
u T^+,
U_0  D_0 (v (T^+ )^{s-1} )\rangle =
\langle
u T^+,
\sum_{n=0}^\infty (D_0 (tD_0 )^n )
(v  (T^+ )^{s-1} )
\rangle\,.
\end{equation}
By partial integration of $D_0$ and by using (\ref{comt}) one gets
\begin{equation}
\label{mbracs1}
\{j_u^2 (\mu ),j_v^s (\mu )\}_\mu =
-\langle
u^\prime T^+,
\sum_{n=1}^\infty (t (D_0 t )^{n-1}D_0 )
(v  (T^+ )^{s-1} )
\rangle -\langle u\tilde{\mu} ,
\sum_{n=1}^\infty ( D_0 (t D_0  )^{n-1} )
(v  (T^+ )^{s-1} )\rangle \,,
\end{equation}
where $\tilde{\mu}= (1-\Pi_+ )\mu$, which implies that the zero grade part of
$\mu$ is cancelled.
In the first term on the right hand side of this equation
we use the skewsymmetric property (\ref{skew}) along with the fact
that $t(T^+ ) =-T^0 $ and
integrate $D_0$ by parts.
In the second term we just
integrate $D_0$ by parts. As a result we obtain
\bee
\label{mbracs2}
\{j_u^2 (\mu ),j_v^s (\mu )\}_\mu &=&
\langle
u^{\prime\prime} T^0,
\sum_{n=1}^\infty ( (tD_0  )^{n-1} )
(v  (T^+ )^{s-1} )
\rangle
-\langle
u^{\prime} [T^0,\mu ],
\sum_{n=1}^\infty ( (tD_0  )^{n-1} )
(v  (T^+ )^{s-1} )
\rangle\nn
&+&\langle ( u\tilde{\mu} )^\prime ,
\sum_{n=1}^\infty  (t D_0  )^{n-1}
(v  (T^+ )^{s-1} )\rangle\,.
\eee

Let us consider the first term on the right hand side of
this equation.
First one observes that the term with $n=1$ vanishes
identically because of the properties of the supertrace.
For the remaining terms we again use the skewsymmetric property
of $t$ and the fact that $t(T^0 )=-\half T^-$ to rewrite them as
\be
\label{center}
\half\langle  u^{\prime\prime}T^-  ,
\sum_{n=2}^\infty  D_0 (t D_0  )^{n-2}
(v  (T^+ )^{s-1} )\rangle = \half \delta (s-2)
      \int dz u v^{\prime\prime\prime} str(T^+ T^- )\,,
\ee
which we recognize as  the central charge in the conformal algebra
(\ref{s2}). To derive this result we used the fact that
the terms with $t$ on the left hand side of (\ref{center})
do not contribute since $\mu$ commutes with $T^-$ and
$str (T^- t (A))=0$ for any $A\in Aq(2;2\nu )$.

For the second term in (\ref{mbracs2}) we first observe
that the commutator $[T^0 ,\mu ]$ is a lowest weight vector.
Hence again all terms containing a $t$ vanish and one is left with
\be
\label{cs}
-\langle
u^{\prime} [T^0,\mu ],
\sum_{n=1}^\infty ( (tD_0  )^{n-1} )
(v  (T^+ )^{s-1} )
\rangle
=
(s-1) \langle
u^{\prime} \mu ,
(v  (T^+ )^{s-1} )
\rangle\,.
\ee
Finally a similar analysis for the third term leads to
\be
\label{3}
\langle ( u\tilde{\mu} )^\prime ,
\sum_{n=1}^\infty  (t D_0  )^{n-1}
(v  (T^+ )^{s-1} )\rangle =
-\langle ( u\tilde{\mu} )        ,
(v^\prime  (T^+ )^{s-1} )\rangle
\,.
\ee

Collecting these three terms together we get
the standard result in the form
(\ref{js})
\be
\label{total}
\{j_u^2 (\mu ),
j_v^s (\mu )\}_\mu =
\int dz ((s-1)u^\prime v -v^\prime u )str\left ((T^+ )^{s-1} \tilde{\mu}
(z)\right ) +\half \delta (s-2) \int dz u v^{\prime\prime\prime} str(T^+ T^-
)\,.
\ee
The appearance of $\tilde{\mu}$ in this formula has the effect that
in the case of the spin - 1 current it indeed has a vanishing
contribution.
One can also directly check that the spin -1 current has the following
Poisson brackets to any spin - $s$ current:
\be
\label{total1}
\{j_u^1 (\mu ),
j_v^s (\mu )\}_\mu =
\delta (s-1) \int dz u v^{\prime} str(1)\,,
\ee
i.e. it decouples from all $s\geq 2$ currents and has only a
central charge contribution in its Poisson bracket with itself.

With similar techniques one can in principle compute any
Poisson bracket between higher-spin currents. It is known
that the full algebrae of higher-spin currents in the Toda
systems are non-linear \cite{conf1}-\cite{conf3},\cite{tw}.
In accordance with the property $(iii)$ of the governing equation
these non-linearities are manifested as non-linearities in $\mu$
since the individual higher-spin currents exhaust all linear
functions of $\mu$. In fact this scheme can be used
in an efficient way to compute $W_n$ algebrae which can be obtained
by specifying $2\nu =2n\pm 1$ (cf (\ref{mattr})). In these cases
all spins higher than $n$ decouple because of the properties of the supertrace.

Taking into account that according to (\ref{js}) $s-1$ coincides
with the negative grade of $\mu$ one can see that any two $j$ parameters
$j_1 (\mu )$ and
$j_2 (\mu )$ with spins $s_1$ and $s_2$, respectively, have some polynomial
Poisson bracket (\ref{j12}). The reason is that each term involving $\mu$ in
$D_0$ will contain an operator $t$ which decreases the grading by one and,
therefore, an order of nonlinearity turns out to be limited by
$min (s_1 ,s_2 )-1$.
Let us note however that there is some difference between the form of
the algebra $\tilde{W}_{1+\infty}$
(i.e. the second Gelfand-Dickey Hamiltonian structure)
discussed in \cite{tw}, which is at most bilinear in fields,
and the
form of the algebra
(\ref{j12}) in which the degree of nonlinearity seems to increase
with spin.
We hope to analyze the relationship between the two algebras in more
details elsewhere.

\setcounter{equation}{0}

\section{Conclusions}

The construction of generalized Toda field theories
performed in Section 3 can be extended also to the
$b(\sigma )$ and $c(\sigma )$ cases by applying the truncations
introduced at the end of Section 2.

The example of the Toda field theory considered in this paper
demonstrates that the infinite-dimensional higher-spin algebras
applied originally to the analysis of the higher-spin gauge dynamics
in two, three and four
space-time dimensions and underlying the model under
consideration can be considered as a natural generalization of the
finite-dimensional matrix algebras. An interesting problem therefore
is to extend the area of applicability of such an approach to
a broader class of models. For example one can think of the generalized
matrix models, etc. An interesting problem is to analyze if it is possible
to extend the
formalism described in this paper to the Affine
Toda theories too.

At the moment we know that there exists a natural
extension of our work to
the supersymmetric case which we will discuss in
a future  publication.
Of key interest is of course also to study
the quantization of these models.

\bigskip
\noindent
{\bf Acknowledgments}
\bigskip

MV is grateful to O.V.~Ogievetsky and M.V.~Saveliev for stimulating
discussions.
LB would like to acknowledge the hospitality of the Lebedev Physical
Institute of the Russian Academy of Sciences.
MV acknowledges that
the research described in this publication
was made possible
in part by Grant \# MQM300 from the International Science Foundation,
grant 93-02-15541 from
the Russian Basic
Research Foundation,
and grant 94-2317 from INTAS.
MV also would like to thank the
Chalmers University for hospitality and
The Royal Swedish Academy of Sciences for the financial support.

\setcounter{section}{1}
\setcounter{equation}{0}
\renewcommand{\thesection}{\Alph{section}}
\bigskip
\noindent

{\Large{\bf{Appendix A}}}
 \vspace{5mm}\\
{\bf Useful Identities}\\ \\

The proof of integrability of higher-spin currents given in Appendix B
is based
on the following important identity
\begin{eqnarray}
\label{cmain}
&{}& G \Big ( A\, G(B)
+ (G(A))\, B\Big )-
G (A)\, G (B)\nonumber\\
&=&
G \Big ((\Pi_- U (A))
G (B)
+G(A)( \Pi_- U (B)\Big )
-V\Big ( \Pi_+\Big ((G(A))\, G(B)\Big )\Big )
\end{eqnarray}
valid for any two
elements  $A$ and $B$ of any arbitrary associative algebra which admits
two derivations $D$ and $T$, i.e.
$D (AB)$=$ D(A) B+AD (B)$ and
$T (AB)$=$ T (A) B+AT (B)$. It is also assumed that there exists a
linear mapping $t$ such that
\be
\label{pg}
Tt=1-\Pi_-\,,\qquad tT=1-\Pi_+
\ee
with some operators $\Pi_\pm$.
The operators $U$ and $G$ are
defined in (\ref{U}) and (\ref{G}).

In applications to the Toda system under consideration
we use the realization (\ref{der}) for $D$ and $[T^+ ,\phantom{M}] $
for $T$.

Let us now prove the identity (\ref{cmain}).
Consider the quantity $X$
\begin{equation}
\label{X}
X=G\Big (T \Big ((G(A))\,(G(B))\Big )\Big )\,.
\end{equation}
By virtue of (\ref{G}) and (\ref{comt}) one gets
\begin{equation}
\label{X1}
X=V\Big ((1-\Pi_+ )\Big ((G(A))\,G(B)\Big )\Big )\,.
\end{equation}
On the other hand applying $T$ to the right in (\ref{X}) and using that
$T$ is a differentiation of the algebra one finds
\begin{equation}
\label{X2}
X=G\{\Big ((1-\Pi_- )\Big (U(A)\Big )\Big )\,(G(B))+
(G(A))\Big ((1-\Pi_- )U(B)\Big )\}\,.
\end{equation}
As a result one arrives at
\bee
\label{I}
&&G\{(U(A))\,(G(B))+(G(A))(U(B))\}-
V\Big (((G(A))\,G(B)\Big )\nonumber\\
&=&
G\{\Pi_- \Big ((U(A))\,G(B)\Big )+(G(A))\Pi_- (U(B))\}
-V\Big (\Pi_+ \Big ((G(A))\,G(B)\Big )\Big )\,.
\eee
By replacing
$U$ by $(U-1)+1$=$ D G +1$ and
$V$ by $(V-1)+1$=$ GD  +1$
with the aid of (\ref{U-1}) and (\ref{V-1}) one transforms the left
hand side of (\ref{I}) to
$$
 G\{(D G(A))\,G(B)+(G(A))(D G(B))\}-
 GD\Big ((G(A))\,G(B)\Big )+
$$
\begin{equation}
G\Big (A\,(G(B))+(G(A))B\Big )-
(G(A))\,G(B)\,.
\end{equation}
Finally one observes that
since $D$ is a differentiation the first three terms in the above
expression cancel so that one is left with the identity (\ref{cmain}).

Let us note that the identity (\ref{cmain})
and the relation (\ref{U-1})
allow us to
identify the following useful new identity
\bee
\label{Uid}
&&U\Big (AG(B)+G(A)B\Big )=U(A)G(B)+G(A)U(B)\nonumber\\
&+&
(U-1)\Big (\Pi_- U(A)G(B) +G(A)\Pi_- U(B)\Big )- D V\Pi_+
(G(A)G(B))\,.
\eee

When the algebra admits a trace operation
$\langle A\rangle$ which induces a
bilinear form $\langle A ,B\rangle  $=
 $\langle A B\rangle  $
 such that
the operators $D$, $T$ and $t$ are skewsymmetric, i.e.
\begin{equation}
\label{asym}
\langle D (A)\rangle =\langle T (A)\rangle =0\,,\qquad
\langle At (B)\rangle =-\langle t (A) B\rangle
\end{equation}
for arbitrary elements $A$ and $B$, the relation (\ref{cmain})
leads to the following one
\be
\label{cmaintr}
\Big \langle\, AG(B)G(C)\, -\Pi_- (U( A))G(B)G(C)\Big \rangle
+\,et\,\, cycl. =0\,.
\ee

Another useful formula which holds for
an arbitrary element $C$ and
 elements
$A$ and $B$ obeying conditions
\be
\label{hv}
[A,B]=0\,,\qquad T (A)=T (B)=0\,
\ee
(these conditions hold e.g. for
highest weight vector elements of $Aq (2;2\nu )$) is
\bee
\label{Y3}
\Big \langle ([V(A), V(B)] C\Big \rangle
&=& \Big \langle [V(A), V(B)] \Pi_-
U(C)\nonumber\\ &+&V(A)[\Pi_- UD(B),G(C)] -V(B)[\Pi_- UD(A),G(C)]\Big
\rangle
\,.
\eee
The proof of this identity is rather technical.
By virtue of (\ref{V-1}) one gets
\bee
\label{Y31}
\Big \langle [V(A), V(B)] C\Big \rangle  =  \Big \langle [GD(A), GD(B)]\,C
+\Big ([GD(A), B] +[A,GD(B) ]\Big )C \Big \rangle
\,.
\eee
Applying (\ref{cmaintr}) to the first term on the right-hand-side of
this relation one arrives at:
\bee
\label{Y32}
\Big \langle [V(A), V(B)] C\Big \rangle &=&
\Big \langle - [D(A), GD(B)]\,G(C)-[GD(A),
D(B)]\,G(C)\nonumber\\
&+&[GD(A),GD(B)]\Pi_- U(C) \nonumber\\
&+& \Big ([\Pi_- UD(A),GD(B)]+[GD(A),\Pi_- UD(B)]\Big )G(C)\nonumber\\
&+&\Big ([GD(A), B] +[A,GD(B) ]\Big )C\Big \rangle \,.
\eee
Now using the simple fact that $D$ is a derivation
one transforms (\ref{Y32}) to
\bee
\label{Y33}
\Big \langle [V(A), V(B)] C\Big \rangle &=&
\Big \langle  - D  \Big ([A, GD(B)]\,G(C)+[GD(A),
B]\,G(C)\Big )\nonumber\\
&+&\Big ([A,DGD(B)]G(C) +[A,GD(B)]DG(C)-A\lr B \Big )
\nonumber\\
&+&[GD(A),GD(B)]\Pi_-U(C) \nonumber\\
&+& \Big ([\Pi_- UD(A),GD(B)]+[GD(A),\Pi_- UD(B)]\Big )G(C)\nonumber\\
&+&\Big ([GD(A), B] +[A,GD(B) ]\Big )C \Big \rangle \,.
\eee
Using (\ref{G}), (\ref{pg}) and (\ref{asym}) one transforms the second line
in the above relation to get
\bee
\label{Y34}
\Big \langle [V(A), V(B)] C\Big \rangle &=&\Big \langle
\Big (AT ([GD(B),G(C)])
- A[D(B),G(C)] -A[GD(B),C]\nonumber\\
&+&A[\Pi_- UD(B),G(C)]) +A[GD(B),\Pi_- U(C)]-A\lr B\Big )
\nonumber\\
&+&[GD(A),GD(B)]\Pi_-U(C) \nonumber\\
&+& \Big ([\Pi_- UD(A),GD(B)]+[GD(A),\Pi_- UD(B)]\Big )G(C)\nonumber\\
&+&\Big ([GD(A), B] +[A,GD(B) ]\Big )C \Big \rangle\,.
\eee
Now one observes that the term involving $T$ vanishes because
of (\ref{hv}) while all terms which do not involve $\Pi_-$
cancel pairwise (taking into account that $[D(A),B]+[A,D(B)]=0$
due to (\ref{hv})). The remaining terms can be easily summed up to
the right hand side of (\ref{Y3}).

In practical calculations it is often useful to use a power series expansion
for the formulae (\ref{cmain}) and (\ref{Y3}) which arises from the geometric
progression formula for $U$ and $V$ in (\ref{U}) and (\ref{V}).

\setcounter{section}{2}
\setcounter{equation}{0}
\renewcommand{\thesection}{\Alph{section}}
\bigskip
\noindent
{\Large{\bf{Appendix B}}}
 \vspace{5mm}\\
{\bf Proofs of the Properties of
the Generalized Currents }\\ \\

In this Appendix we
 prove the properties (i)-(vi) of the generalized currents
formulated in Section 6.

Proof for Proposition (i):

A straightforward calculation shows
$$
\{{\cal H},J\}_\phi \vert_{\mu =0} =\langle T^+ ,[H^- ,\frac{\delta J}
{\delta \phi^\prime}]\rangle \vert_{\mu =0} = -\langle H^- ,[T^+ ,
\frac{\delta J}
{\delta \phi^\prime}]\rangle \vert_{\mu =0}\,.
$$
By virtue of the governing equation (\ref{gov}) we rewrite this
expression as
$$
\langle \eta , \frac{\delta J}
{\delta \mu}\rangle \vert_{\mu =0}
$$
with
$$
\eta =-\Pi_- U ([T^+ ,H^- ]) \vert_{\mu =0}\,.
$$
Next we use (\ref{U-1}) and (\ref{comt}) to rewrite $\eta$ as
$$
\eta =-\Pi_- \Big ([T^+ ,H^- ] + {\cal U} ({\cal D} (1-\Pi_+ ) H^- )\Big )\,.
$$
One then observes that ${\cal D} H^- \equiv 0$
(${\cal D} H^- =R_{+-}^- \equiv 0$ due
to the definition of the Lorentz connection in the original zero curvature
 approach), $\Pi_+  H^- =0$ and $\Pi_- ([T^+ , H^- ] )=0$.  This concludes the
proof of Proposition (i).

The proofs of the Propositions (ii) and (iii) are obvious.

Proof for Proposition (iv):

In order to prove that $\Dc_\xi
(\langle \frac{\delta J_{1}}{\delta \mu},
UD(\frac{\delta J_{2}}{\delta \mu})\rangle )=0$ provided that
$\Dc_\xi J_{1}$=$\Dc_\xi J_{2}=0$ consider the following expression
\be
\label{41}
X=\Dc_\xi \Big (\langle \frac{\delta J_{2}}{\delta \mu},
UD(\frac{\delta J_{1}}{\delta \mu})\rangle \Big ) -
 \langle \frac{\delta \Dc_\xi (J_{2})}{\delta \mu},
UD(\frac{\delta J_{1}}{\delta \mu})\rangle  -
 \langle \frac{\delta J_{2}}{\delta \mu},
UD(\frac{\delta \Dc_\xi (J_{1})}{\delta \mu})\rangle\,.
\ee
Using the explicit forms of the $\Dc_\xi $ and $D$ one finds
\be
\label{42}
X=
\langle \frac{\delta J_{2}}{\delta \mu},
U[\xi -\Pi_- U(\xi ),V(
\frac{\delta J_{1}}{\delta \mu})]\rangle
+
\Big (\langle \frac{\delta J_{2}}{\delta \mu},
U[\Pi_- UD\frac{\delta J_{1}}{\delta \mu},tU(\xi)]\rangle -(1\lr 2 )\Big )\,.
\ee
By identifying
$\frac{\delta J_{2}}{\delta \mu}$,
$\frac{\delta J_{1}}{\delta \mu}$
and $\xi$ with $A$, $B$, and $C$
respectively one concludes that $X=0$ as a consequence of
the identity (\ref{Y3}).
This concludes the proof of Proposition (iv).

Let us now explain why proposition (v) is true. {}From the definition of the
Poisson structure (\ref{shift}) one gets
$$ \{J_1 ,J_2 \}_\phi = \langle
\frac{\delta J_1}{\delta\phi^\prime} ,\partial_z \frac{\delta
J_2}{\delta\phi^\prime} \rangle \,.
$$
By virtue of the governing equation (\ref{gov})
one finds
$$ \{J_1 ,J_2 \}_\phi =  \langle \frac{\delta J_1}{\delta\mu }
,U\partial_z \frac{\delta J_2}{\delta\phi^\prime} \rangle \,.
$$ Using the
governing equation a second time one gets
$$
\{J_1 ,J_2 \}_\phi =-
\langle \frac{\delta J_2}{\delta\mu } , U\Pi_0 \partial_z V
\frac{\delta J_1}{\delta\mu} \rangle \,.
$$
The crucial observation now is that when $\mu=0$ one can
first replace $\partial_z$ by $\D$ in the above formula
(because of the operator $\Pi_0\,$, the term
in $\D$ which involves the commutator vanishes since
$\phi^\prime$ commutes with neutral elements) and,
second, use that
\be
\label{deco}
\langle \rho_1 , U\Pi_0 \partial_z
V\rho_2\rangle\vert_{\mu =0} =
\langle\rho_1 , UD\rho_2 \rangle \vert_{\mu =0}
\ee
for arbitrary positive graded $\rho_{1,2}$ because
$D\vert_{\mu =0} =\D$ has grading zero while $t$ has grading $-1$,
so that the left hand side just accounts for all possibilities for
insertions of the projection operator $\Pi_0$
inside those terms
on the right hand side
which have a non-vanishing supertrace.
This completes the proof of Proposition (iv).

To prove the proposition (vi) one observes that it follows from (\ref{j12})
that
\bee
\label{61}
X_{1,2,3}&=&\{\{j_1 ,j_2 \}_\mu ,j_3 \}_\mu +\ec
=\langle V_0(A_1 ),[\Pi_- (D_0 V_0 (A_3 )),V_0 (A_2 )]\rangle +\ec\nn &=&
-\langle [V_0 (A_1 ), V_0 (A_2 )],\Pi_- D_0 V_0 (A_3 )\rangle\,+\ec \,,
\eee
where we use the notation
$A_i =\frac{\delta j_i}{\delta \mu}$
so that $A_i$ are highest weight vectors obeying the
properties (\ref{hv}).
With the aid of the substitution (\ref{V-1})
one gets
\bee
\label{63}
X_{1,2,3}
&=&
-\Big \langle
[G_0 D_0 (A_1 ), G_0 D_0 (A_2 )]\Pi_- D_0 V_0 (A_3 )\Big \rangle\nonumber\\
&-&\Big \langle\Big ([A_1 ,G_0 D_0  (A_2 )] +[G_0 D_0 (A_1 ), A_2 ]\Big ) \Pi_
- D_0 V_0 (A_3  )
\Big \rangle +\ec \,.
\eee
Then, using the basic identity
(\ref{cmaintr}) for the first term on the right hand side of (\ref{63})
one finds
\bee
\label{64}
X_{1,2,3}
&=&
-\langle D_0 (A_1 ) [G_0 D_0 (A_2 ),G_0 D_0 (A_3 )]\rangle\nn
&-&\langle ([A_1 ,G_0 D_0  (A_2 )] +[G_0 D_0 (A_1 ), A_2 ])\Pi_- D_0 V_0 (A_3 )
\rangle+\ec \,.
\eee
Using (\ref{V-1}) once more along with the cyclic property of
the supertrace and the fact that $[A_i ,A_j ]=0$ and therefore
$[DA_i ,A_j ]+[A_i ,DA_j ]=0$
one finds
\bee
\label{65}
X_{1,2,3}
=
&-&\langle D_0 (A_1 ) [V_0 (A_2 ),V_0 (A_3 )]\rangle\nn
&-&\langle ([A_1 ,V_0 (A_2 )] +[V_0 (A_1 ), A_2 ])\Pi_- D_0 V_0 (A_3 )
\rangle +\ec\,.
\eee
By an  integration by parts
\bee
\label{66}
X_{1,2,3}
&=&
\langle A_1  ([D_0 V_0 (A_2 ),V_0 (A_3 )]+[V_0 (A_2 ),D_0 V_0 (A_3 )]\rangle\nn
&-&\langle ([A_1 ,V_0 (A_2 )] +[V_0 (A_1 ), A_2 ])\Pi_- D_0 V_0 (A_3 )
\rangle +\ec\,.
\eee
Next we use the following identity
\bee
\label{67}
&0&=\langle A_1 ,[T^+ ,[V_0 (A_2 ),V_0 (A_3 )]]\rangle +\ec
\nonumber\\&=&
\langle A_1  ([(1-\Pi_- )D_0 V_0 (A_2 ),V_0 (A_3 )]
+[V_0 (A_2 ),(1-\Pi_- )D_0 V_0 (A_3)]\rangle +\ec ,
\eee
which is a consequence of
$\langle [A_i ,[T^+ ,B]]\rangle$
= $\langle [[A_i ,T^+ ],B]\rangle =0$ and (\ref{comt}), (\ref{G}).
Finally one observes that the right-hand side of (\ref{66})
exactly coincides with
the expression on the right-hand side of (\ref{67}).
This completes the proof that $X_{1,2,3}=0$.

\newpage
\setcounter{section}{3}
\setcounter{equation}{0}
\renewcommand{\thesection}{\Alph{section}}
\bigskip
\noindent
{\Large{\bf{Appendix C}}}
 \vspace{5mm}\\
{\bf Notations and Definitions }\\ \\

\be
T^0 = \frac{1}{4} \{a^- , a^+ \}\,,\qquad
T^\pm = \frac{1}{2} (a^\pm )^2\,,
\nonumber
\end{equation}
\begin{equation}
[T^0 ,T^\pm ] =\pm T^\pm\,,\qquad [T^- ,T^+ ]=2T^0\,,
\nonumber
\end{equation}
\be
E_{n,m}=\Big ( (a^+ )^n (a^- )^m \Big )_W \equiv\frac{1}{(n+m)!}
\Bigl ( (a^+ )^n (a^- )^m +((n+m)!-1)  )\,\mbox{permutations}\Bigr ) \,,
\nonumber\ee
\be
h_n =E_{n,n}\,,
\nonumber\ee
\bee
str(P^\pm E_{n,m}E_{m,n}) =(-1)^n n!\,m!\beta^\pm (n+m)\,,
\nonumber
\eee
\be
\partial_\pm=\frac{1}{\sqrt{2}}(\partial_0 \pm \partial_1 )\,,
\nonumber\ee
\begin{equation}
{\cal D} A=\partial_+ A + [\partial_+ \phi , A]\,,
\nonumber
\end{equation}
\begin{equation}
t(E_{n,m})=-{1\over m+1}E_{n-1,m+1}\quad  n>0\,;\qquad  t(E_{0,m})=0\,,
\nonumber
\end{equation}
\begin{equation}
[T^+ ,t(x)] =x-\Pi_- (x) \,,\qquad
t([T^+ ,x]) =x-\Pi_+ (x)\,,
\nonumber
\end{equation}
\begin{equation}
\Pi_- (E_{0,m})=E_{0,m}\,;\qquad
\Pi_- (E_{n,m})=0\quad n>0\,,
\nonumber
\end{equation}
\begin{equation}
\Pi_+ (E_{n,0})=E_{n,0}\,;\qquad
\Pi_+ (E_{n,m})=0\quad m>0\,,
\nonumber
\end{equation}
\begin{equation}
\Pi_0 (E_{n,n})=E_{n,n}\equiv h_n\,;\qquad
\Pi_0 (E_{n,m})=0\quad n\ne m\,,
\nonumber
\end{equation}
\begin{equation}
\langle\xi , \phi \rangle =\int dz str \xi (z) \phi (z)\,,
\nonumber
\end{equation}
\begin{equation}
\phi^\prime = \partial_+ \phi\,.
\end{equation}
\begin{equation}
\langle\xi^\prime ,\eta \rangle +
\langle\xi ,\eta^\prime \rangle =0\,,
\nonumber
\end{equation}
\begin{equation}
\{\langle\xi , \phi \rangle ,\langle\eta , \phi^\prime \rangle\}=
\langle\xi ,\eta \rangle\,,
\nonumber
\end{equation}
\begin{equation}
\Dc_\xi =
\langle\xi ,
\frac{\delta}{\delta\phi^\prime} \rangle
- \langle \Pi_- U \xi ,
\frac{\delta}{\delta\mu}
 \rangle\,,
\nonumber
\end{equation}
\begin{equation}
U=(1- D t)^{-1}\,,
\end{equation}
\begin{equation}
D(X)=\partial_z (X)+ [\phi^\prime ,X] +[\mu ,X]
\nonumber
\end{equation}
\be
\langle \xi ,
\frac{\delta}{\delta\phi^\prime}\rangle \phi^\prime  =\Pi_0 (\xi )\,;\qquad
\langle \eta ,
\frac{\delta}{\delta\mu}\rangle \mu  =\Pi_- (\eta ) \,.
\nonumber\ee
\begin{equation}
V=(1- tD )^{-1}\,,
\nonumber\end{equation}
\begin{equation}
G=tU=Vt\,.
\nonumber\end{equation}
\begin{equation}
j_{1, 2}(\mu )=\langle \frac{\delta j_{1}}{\delta \mu},
U_0  D_0 (\frac{\delta j_{2}}{\delta \mu})\rangle\,,
\nonumber\end{equation}
\begin{equation}
U_0 =(1- D_0 t)^{-1}\,,\qquad
V_0 =(1- t D_0 )^{-1}\,,
\nonumber\end{equation}
\be
{\cal U}= (1-\D t )^{-1}\,,\qquad \D =\partial_z +[\phi^\prime ,
\phantom{M}]
\,,
\nonumber
\ee
\be
{\cal V}= (1-t \D  )^{-1}\,.
\nonumber
\ee

\end{document}